\documentclass[12pt]{iopart}

\expandafter\let\csname equation*\endcsname\relax
\expandafter\let\csname endequation*\endcsname\relax

\usepackage{amsmath}

\usepackage{adjustbox}
\usepackage[linesnumbered,ruled,vlined]{algorithm2e}
\usepackage{algpseudocode}
\usepackage[colorlinks=true,urlcolor=blue,citecolor=blue,linkcolor=blue]{hyperref}
\usepackage{graphicx}
\usepackage{caption}
\usepackage{color}
\usepackage{bm}
\usepackage{dsfont}
\usepackage{amssymb}
\usepackage{iopams} 
\usepackage{bm}
\usepackage[page,header]{appendix}
\usepackage{titletoc}

\usepackage[all]{nowidow}
\usepackage{quantikz}
\usepackage{tikz}
\usepackage{thmtools}
\usepackage{multirow}

\graphicspath{{Figures/}}

\begin{document}

\title{Prospects for Quantum Computation in  Propellant Design: Assessing the Stability  of Cyclic Ozone in Nanoscale Confinement}

\author{Thomas W. Watts$^1$, Matthew Otten$^2$, Jason T. Necaise$^3$, Nam Nguyen$^4$, Benjamin Link$^5$, Kristen S. Williams$^6$, Yuval R. Sanders$^7$, Samuel J. Elman$^7$, Maria Kieferova$^7$, Michael J. Bremner$^{7,8}$, Kaitlyn J. Morrell$^9$, Justin E. Elenewski$^9$, Samuel D. Johnson$^1$, Luke Mathieson$^7$, Kevin M. Obenland$^9$, Rashmi Sundareswara$^1$, Adam Holmes$^1$}

\address{$^1$ HRL Laboratories, LLC, Malibu, CA, USA}
\address{$^2$ Department of Physics, University of Wisconsin -- Madison, Madison, WI, USA}
\address{$^3$ Dartmouth College, Hanover, NH, USA}
\address{$^4$ Applied Mathematics, Boeing Research \& Technology, Huntington Beach, CA, USA}
\address{$^5$ Applied Mathematics, Boeing Research \& Technology, Orlando, FL, USA}
\address{$^6$ Applied Mathematics, Boeing Research \& Technology, Huntsville, AL, USA}
\address{$^7$ Centre for Quantum Software and Information, School of Computer Science, Faculty of Engineering \& Information Technology, University of Technology Sydney, NSW 2007, Australia}
\address{$^8$ Centre for Quantum Computation and Communication Technology, University of Technology Sydney, NSW 2007, Australia}
\address{$^9$ MIT Lincoln Laboratory, Lexington, MA, USA}

\ead{thomas.watts@student.uts.edu.au}

\begin{abstract}
Cyclic ozone additives have the potential to markedly increase the specific impulse of rocket fuel. This would translate to greater efficiency and reduced costs for space lift by granting more payload per rocket.  While  practical efforts to capture this isomer have been unsuccessful, it is possible that cyclic ozone would be stabilized in confined geometries. The required synthetic methods are nonetheless difficult to design and require theory--driven inputs that lie beyond the scope of classical methods.  Quantum computation has the potential to enable these calculations, though the underlying hardware requirements remain unclear for many practical applications. We present an end-to-end analysis of how quantum methods could support   efforts to isolate cyclic ozone via fullerene encapsulation.  Our discussion extends beyond asymptotic complexity, reporting both logical- and physical-level resource estimates for ground-state energy determination via quantum phase estimation (QPE), computed using multiple independent resource estimation stacks—Azure Quantum Resource Estimator (AzureQRE), PennyLane’s resource estimation framework, and MIT Lincoln Lab's pyLIQTR toolkit—to cross-check resource estimates and bracket realistic overheads. Taken collectively, these data delineate a plausible scale for realistic, computationally--aided molecular design efforts using fault--tolerant quantum computation.
\end{abstract}

\maketitle
\tableofcontents
\setcounter{tocdepth}{1}
\newpage
\section{Introduction}
\label{section:Introduction}

\subsection{Overview}

\begin{figure}
\begin{center}
\includegraphics[width=\textwidth]{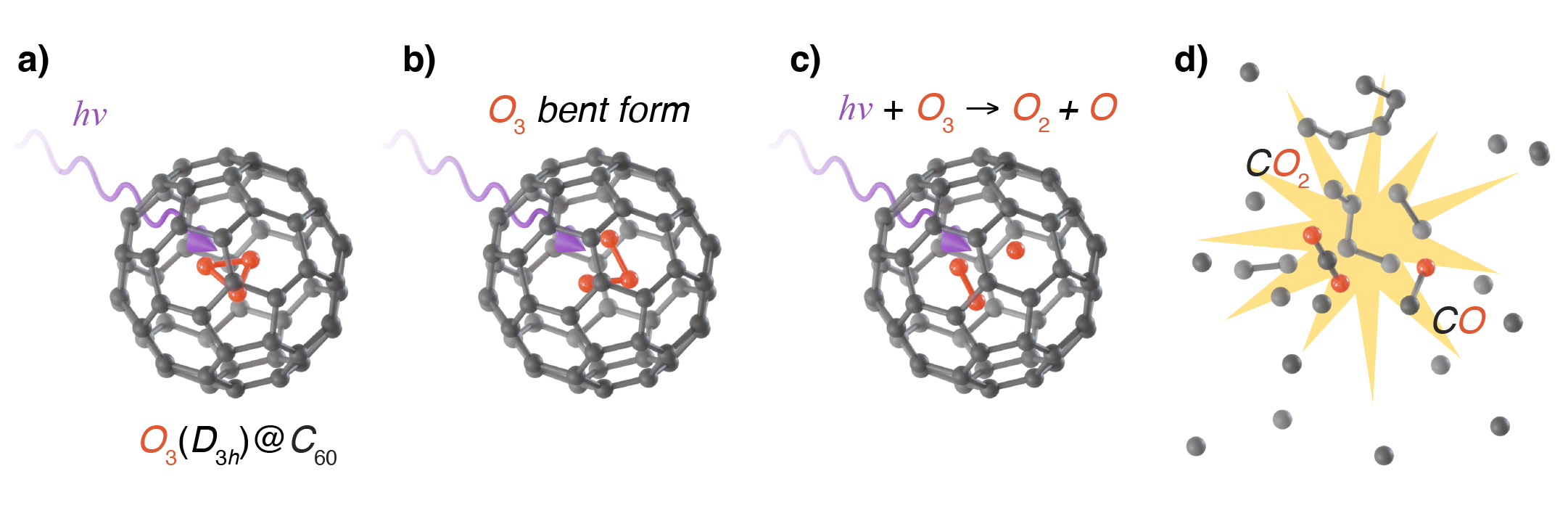}
\caption{{\bf Proposed photolysis of fullerene--encapsulated cyclic ozone.}   (a) Cyclic ozone O$_3(D_{3h})$ is isolated in a fullerene matrix, depicted here as C$_{60}$; (b) single--photon excitation photoisomerizes  O$_3(D_{3h})$  to the conventional bent isomer  O$_3(C_{2v})$, inducing molecular strain in already tight geometries;  (c) bent ozone undergoes subsequent photolysis into molecular O$_2$ and atomic oxygen O; (d) the resulting species react with the fullerene cage through combustion reactions, and foster a sustained burn when numerous oxygen equivalents are present.  }
\label{fig:cage_burst}
\end{center}
\end{figure}

The development of advanced rocket fuels is critical to the  efficiency and performance of future space missions.  While fundamentally effective, traditional chemical propellants are often limited in terms of energy density, stability, and environmental impact.  These parameters must be balanced when selecting a fuel, and this competition has excluded otherwise appealing candidates.  For instance, ozone O$_3$  has a very favorable energy density, but this is counterbalanced by its excessive reactivity~\cite{Chen2011}.  Ozone's cyclic isomer O$_3(D_{3h})$ \footnote{We will differentiate bent O$_3(C_{2v})$ and cyclic O$_3(D_{3h})$ isomers of ozone by appending the Schoenflies symbol for the latter. The symmetry group will be omitted for the bent form by convention.} is predicted to carry an even higher energy density, though it has not been isolated owing to commensurate instability~\cite{laser_search_article, o3_laser_pulse_search, mgo_crystal_ozone}.

\par Beyond conventional molecular propellants, a broad class of \emph{nanoenergetic} materials leverages nanoscale structuring to tune reaction rates, ignition thresholds, and heat/pressure release in compact chemical systems~\cite{Trache2020Nanoenergetic,Thiruvengadathan2024Nanoenergetic,Rossi2007MEMS,Martirosyan2011GasGenerators}. A recurring theme is that confinement and interfacial control at the nanometer scale can qualitatively modify transport and reactivity, enabling rapid energy release in a small footprint and offering new design knobs for initiation and stability~\cite{Trache2020Nanoenergetic,Thiruvengadathan2024Nanoenergetic}. In parallel, endohedral fullerenes provide a chemically robust \emph{nanocontainer} platform in which encapsulation can perturb electronic structure and reaction landscapes, motivating their use as model confined reactors for reactive species~\cite{Popov2013Endohedral,Jalife2020NgEndohedral}. Nanoenergetic materials enable an ecosystem of microsystem technologies: chip-scale initiators, rapid micro-actuators, and compact gas-generators that can deliver a sharp, programmable impulse on demand~\cite{Rossi2007MEMS,Martirosyan2011GasGenerators}. In aerospace and small-satellite contexts, related micropropulsion hardware (e.g., MEMS solid microthruster arrays) has been actively reviewed as a route to low-mass, fast-response attitude control and small maneuver capability, where integration, reliability, and reproducibility matter as much as peak performance~\cite{Xu2024MicrothrusterReview}.

\par Recent theoretical efforts suggest that cyclic ozone might be synthesized and stabilized within a fullerene matrix, opening one avenue toward its practical use~\cite{entropy_of_ozone, Sharma2025}. This assumption is not without precedent, since geometric confinement is known to shift the relative energetics of different molecular isomers and conformers~\cite{entropy_of_ozone, Sharma2025, nature2025, Saunders1993}.  Moreover, it has been speculated that cyclic H$_3$ could be stabilized in this manner for hydrogen storage~\cite{hydrogen_storage}.  Cyclic ozone may also have unique features as a propellant, including photochemically initiated combustion.  As depicted in Fig.~\ref{fig:cage_burst}, ultraviolet light could penetrate a fullerene cage, generating heat and reactive oxygen species through photolysis~\cite{patent_uv_synth}.  Although fullerenes have a nontrivial UV absorption profile --- limiting the ultimate optical path depth in a medium --- it would still be possible to initiate a vigorously exothermic burn from the boundary. The prospects for controlled energy release, large--scale storage, and safe handling are desirable features in an advanced propellant~\cite{Kroto1985,Saunders1993}.

\par While stabilized cyclic ozone is appealing, its synthesis remains a formidable challenge.  The leading strategy for fullerene encapsulation leverages a procedure termed \textit{fullerene surgery}.  Here, an existing fullerene is opened via cycloaddition reactions or electrocylic shifts, while leaving a synthetic handle to reseal the cage after introducing a small molecule ~\cite{fullerene_surgery, synthesis_of_endofullerenes}.  The resulting arrangement is termed an endohedral fullerene or endofullerene.  This route for encapsulation was proposed in a recent patent for sunscreen, textiles, sunglasses, and wind screens~\cite{patent_uv_synth}.  More specifically, this strategy would  sequester  oxygen molecules (O$_2$) in fullerenes of varying size and bombard them  with oxygen ions to generate ozone.   The stability of any products would be dictated by the size of the fullerene cage and the number of encapsulated oxygen equivalents.  By tuning these factors, it may be possible to shift thermodynamic equilibria in favor of ozone's cyclic isomer.  Alternatively, optical excitation could be used to generate cyclic ozone \emph{in situ} from its bent counterpart.  However, these proposals assume that cyclic ozone would be stable in a fullerene matrix.

\par
\begin{figure}
    \centering
    \includegraphics[width=0.4\linewidth]{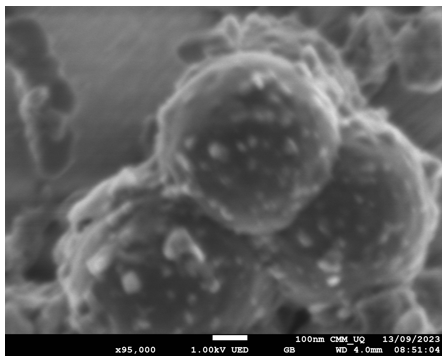}
    \caption{{\bf SEM image of carbon nano-onion (CNO) molecules courtesy of Advanced Nanomaterials \cite{AdvancedNanomaterials}.}}
    \label{fig:cno}
\end{figure}

 Fullerene surgery is a very expensive process that produces only grams of endofullerene molecules  practically. New methods for synthesizing complex endofullerenes in bulk have recently been developed. Advanced Nanomaterials uses a manufacturing process that mimics star-like conditions to produce large quantities of a complex mix of nanocarbon materials, including carbon nano onions (CNO), endofullerenes, and other complex nanocarbon allotropes \cite{AdvancedNanomaterials}. Fig.~\ref{fig:cno} provides an SEM image of CNOs synthesized by Advanced Nanomaterials. The manufacturing equipment and process took thousands of experiments to perfect. Their processes involve removing as much oxygen from their crucibles as possible, given its explosive nature. Adapting their methods for the synthesis of oxygen-containing endofullerenes should be carried out with extreme caution.

\par
\begin{figure}[t]
\centering
\includegraphics[width=\textwidth]{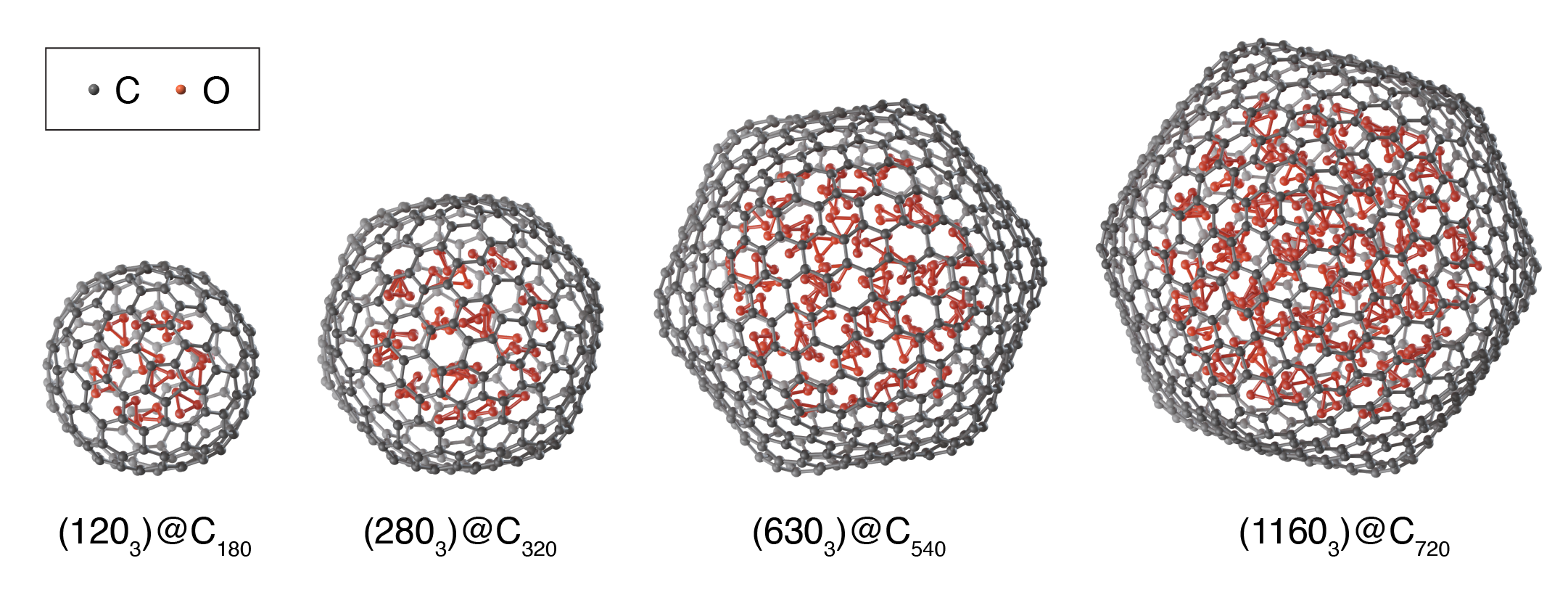}
\caption{ {\bf Relaxed geometries of ozone--containing endofullerenes.} Classically optimized geometries of ($x\text{ O}_3(D_{3h}))$@C$_n$ for $(x,n) = (12,180), (28,320), (63,540), (116,720)$, corresponding to utility--scale problem instances that might be solved with quantum computation.  Relaxed geometries were obtained using density functional theory (DFT), where electronic structure is described using a PBE+D3 scheme \cite{Perdew1996, Grimme2010} with a GTH pseudopotentials \cite{Goedecker1996} and the TZVP basis set \cite{Schaefer1994}.  Calculations were facilitated by the CP2K code \cite{cp2k}.}
\label{fig:endofullerenes}
\end{figure}

 The complexity of these experiments would preclude a systematic, trial--and--error strategy.  At the minimum,  constraints on the target fullerene size and oxygen loading fraction should be known from the outset. An application--oriented perspective would also require some knowledge about reaction mechanism -- including photochemical initiation or synthesis -- and thus capture prospects for use as a rocket propellant. Accurate reaction energies are particularly relevant when predicting rate constants $k \propto \exp[-\Delta E_a / kT]$, which scale exponentially in ratio of activation energy $\Delta E_a$  to the thermal energy scale $kT$.  Classical quantum chemical methods formally offer a means to assess the relative energetics of endofullerenes ($x\text{ O}_2$, $y \text{ O}_3(D_{3h})$, $z \text{ O}_3(C_{2v})$)@C$_n$ and their reactive pathways, where $(x,y,z)$ specify the stoichiometry of oxygen species and $n$ the size of the fullerene (Fig.~\ref{fig:endofullerenes}).  Unfortunately, ozone's electronic structure and reactivity are characterized by nontrivial non--adiabatic effects and pronounced multireference character~\cite{casscf_ozone_study}\footnote{The term multireference is used in the chemistry community to describe an electronic structure that isn't captured by a single Slater determinant (e.g., as might be obtained from a Hartree--Fock calculation)~\cite{Buenker1974,Buenker1978,Werner1988}. Instead, they require a linear combination of multiple reference states in order to fully resolve the strong correlation present in the electronic structure.}.  This means that most cost--efficient classical computational methods will be inapplicable. While the ground state energy of a single isolated ozone molecule has been computed to high accuracy via the semistochastic heat--bath configuration interaction (SHCI) method, this technique is insurmountably expensive for the minimal example of O$_3$@C$_{60}$ ~\cite{SHCI_ozone_pes}. 

\par The electronic structure problem is notoriously difficult due to the exponential scaling between Hilbert space dimension and problem size (e.g., electron filling and orbital count)~\cite{McArdle2020}. This translates to a commensurate overhead in terms of classical resources such as memory and runtime. Quantum computation has the potential to reduce this cost using resources that are  polynomial in problem scale.  Indeed, quantum chemistry is believed to be one of the first applications that will see a fruitful return from the use of quantum algorithms \cite{Ortiz2000QuantumAF, Kassal2008PolynomialtimeQA, Lanyon2009TowardsQC, chem_rxn_qc, Bauer_2020}.  These prospects have advanced rapidly, resulting in quantum algorithms with reduced resource overhead and the ability to treat more diverse processes such as time--dependent dynamics \cite{Low2019hamiltonian, Low2018HamiltonianSI, hypercontraction, quantum_materials_simulation, linear_t_complexity_dpw, Berry2024DoublingEO}.

\par In this contribution, we estimate the computational cost of solving the electronic structure of fullerene--encapsulated cyclic ozone using fault--tolerant quantum computation.  Our perspective on quantum computation assumes state--of--the--art algorithms, including qubitized  phase estimation (QPE) and efficient block encodings for the electronic structure Hamiltonian.  This arrangement would provide energy determinations comparable to FCI in accuracy, yet tackle systems that are intractable through contemporary classical means.  We capture this in a concrete setting, adopting the (photo)isomerization reactions O$_3(D_{3h})$ $\rightleftarrows$ O$_3(C_{2v})$ as a target.  A clear perspective on this process would be needed in order to capture thermal stability or the photochemically--aided synthesis or ignition routes.  However, the constituent estimates also reflect the scale of other calculations that predict stability.  These estimates collectively assess a realistic quantum computational workflow for a practical problem in molecular engineering. Our computational workflow can be extended to a variety of confined geometries such as molecules encapsulated in doped-fullerenes or boron-based cages. A very interest avenue for future research would be to see if fullerenes can stabilize larger allotropes of oxygen that have attractive entropies $h \in [0,1]$, namely O$_8(D_{4h})$ which is the only other oxygen allotrope with $h=0$ besides the diatomic O$_2(D_{\infty h})$ and cyclic ozone O$_3(D_{3h})$ \cite{entropy_of_ozone}.




\subsection{Quantum Computational Approach}
\label{subsec: quantum amenability}
\par To understand the feasibility of constructing complex macromolecules --- such as endohedral fullerenes --- we need to capture the energetics of their reaction pathways.  These data are also used to infer stability measures and to assess the viability of  rocket propellants.  Quantum computing may offer a resource efficient means to handle systems such as ozone, which are characterized by strong multi--reference character and non--adiabatic dynamics. The quantum phase estimation (QPE) algorithm is central to these efforts, and can capture state energies of an electronic structure Hamiltonian with sub--exponential overhead in system size~\cite{Gharibian2022}.  This is markedly more efficient than comparable classical methods such as the full configuration interaction (FCI), which invariably has exponential overhead and is limited to small systems. The ability to treat large systems the FCI level will translate to increased accuracy (e.g., since QPE could be chosen over approximate classical methods, which are insufficient for this system).   There is an important caveat, in that QPE can formally require exponential resources to yield high-precision estimates of ground--state energies with high probability.  However, this assumes a poor ansatz for the initial state and tight error thresholds.  The problem we consider is highly structured and we do not expect its complexity to scale like the general case.

\begin{figure}[t]
    \centering
    \includegraphics[scale=1.1]{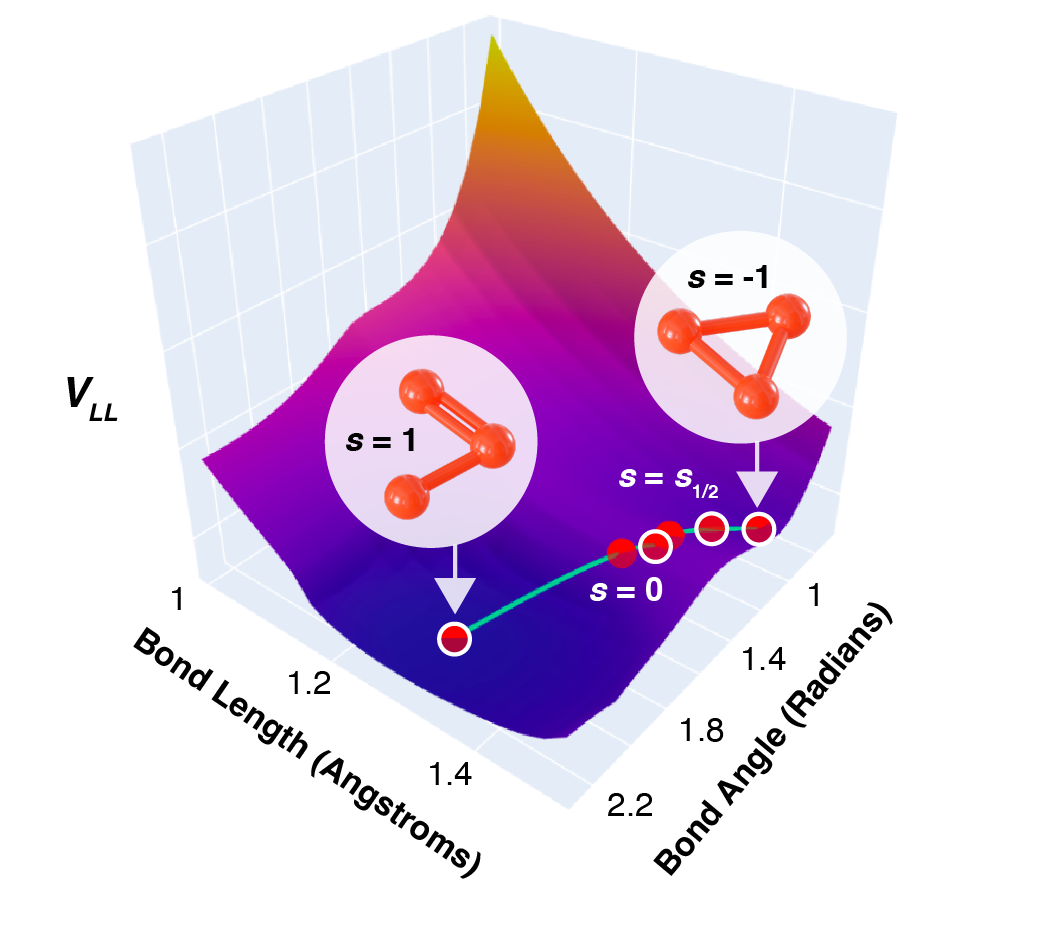}
    \caption{ {\bf Potential energy surface for ozone isomerization}. The $1 {}^1{A}_1$  ground--state potential energy surface for ozone isomerization is presented at the PBE/GTH/TZVP level, with the minimum energy path (reaction coordinate) between bent ($s=1$) and cyclic ($s=-1$) isomers depicted as a green line. These data correspond to the low--level contribution $V_{\text{LL}}$  from the VTST formalism. The four red points with white circles correspond to the thermodynamic endpoints ($s= \pm 1$) as well as the transition state ($s=0$) and an additional point used for fitting ($s=s_{1/2}$).  The remaining red points capture crossing between the $1 {}^1{A}_1$ ground state  and the $1 {}^1{A}_2$ excited state surfaces along the reaction coordinate.  Computational methods are identical to those used in Fig.~\ref{fig:endofullerenes}.}
    \label{fig:low_level_pes}
\end{figure}

\par Our analysis aims to capture the thermal or photo--isomerization  between bent and cyclic isomers of an ozone molecule O$_3(C_{2v}) \leftrightarrows $ O$_3(D_{3h})$ that is sequestered in an endofullerene with other oxygen equivalents.  Some isomerization reactions,  such as those addressed in our efforts, require a higher level of theory due to subtle energetic differences between participating species or contributions from nuclear quantum effects.  Indeed, prior  computational studies of this reaction suggest oxygen tunneling may play an important role in the thermal isomerization rate~\cite{thermal_stability_of_cyclic_ozone}.

\par In the most general case, QPE could be used to map a full potential energy surface (PES) for the isomerization reaction.  This would involve many calculations with significant quantum resource overhead and overestimates the cost of a utility--scale task.  Substantial utility can also be found in augmenting classical calculations, which would require far fewer quantum datapoints.  We imagine a kinetic estimation workflow based on variational transition state theory (VTST; Fig.~\ref{fig:low_level_pes}), which uses a correction function $\Delta V(s) = V_{\text{HL}}(s) - V_{\text{LL}}(s)$ to reconcile low--accuracy energies $V_{\text{LL}}(s)$ along a reaction coordinate $s$ with counterparts $V_{\text{HL}}(s)$ from a higher level of theory.   To give a concrete example, the contributions $V_{\text{LL}}(s)$ may be taken from density functional theory while $V_{\text{HL}}(s)$ would come from a quantum QPE calculation.  We may not require $V_{\text{HL}}(s)$ for every point along $s$, as interpolation may be sufficient for certain segments. Moreover, some rate--estimation methodologies will only use a subset of these points.  VTST and related methods have been quite fruitful for isomerization reactions that involve conical intersections and tunneling phenomena~\cite{vtst_review_paper, VTST_for_rxn_pathway}. 

We adapt an approach that previously captured the thermal stability of an isolated cyclic ozone molecule~\cite{thermal_stability_of_cyclic_ozone}. Following this method, we require at least four FCI--quality ground state energies $V_{\text{HL}}(s)$ along the reaction coordinate.  These are taken at the endpoints $s=\pm 1$ where $s=\pm 1$ are reactant (cyclic ozone, $s=-1$) and product (bent ozone, $s=1$), and at the transition state $s=0$.  We also include an auxiliary point $s_{1/2} \in [-1,0)$ that helps determine the correct fitting parameters for the correction function $\Delta V$.  This point should not be too close to the transition state.  The thermal isomerization pathway for ozone along a DFT--defined PES is depicted in Fig.~\ref{fig:low_level_pes}, and the same conceptual arrangement would be used for photoisomerization on an excited state surface (e.g., following vertical excitation from the ground state surface). We estimate the quantum computational cost of producing FCI--quality energies at points $s=\pm 1, 0, s_{1/2}$ using QPE, assuming that a single ozone isomer is being tracked in the endofullerene $(x$ O$_2 ,y$ O$_3)@$C$_n$.  While several structural configurations would likely be sampled in a practical workflow (e.g., such as snapshots from an \emph{ab initio} dynamics trajectory), this gives only a multiplicative factor on top of our estimates.

\section{Workflow and quantum algorithm framework}\label{sec:workflow}

\subsection{Computational Workflow}

\par Our target is the isomerization rate constant $k(T)$ between cyclic and bent ozone at a fixed temperature $T$.  We assume that this occurs within a fullerene and in the company of other molecular oxygen equivalents:	

\begin{equation}
(x\text{ O}_3(D_{3h}), y\text{ O}_2) @\text{C}_n \xrightarrow{k(T)} (\text{O}_3(C_{2v}), (x-1)\text{ O}_3(D_{3h}), y\text{ O}_2) @\text{C}_n.
\end{equation}

\noindent This reaction underlies the thermal decomposition or photochemical activation of cyclic ozone in a fuel mixture. However, the same computational strategy would apply to reverse reaction, which synthesizes of cyclic ozone from its bent counterpart.  Note that structural changes from isomerization can also induce strain the fullerene system, ultimately leading to C--C bond scission and opening of the fullerene cage.  This possibility has already been explored in the context of hydrogen storage~\cite{hydrogen_storage} and could be accommodated in a more comprehensive treatment.

\par
\begin{figure}[t]
\begin{center}
    \includegraphics[scale=0.55]{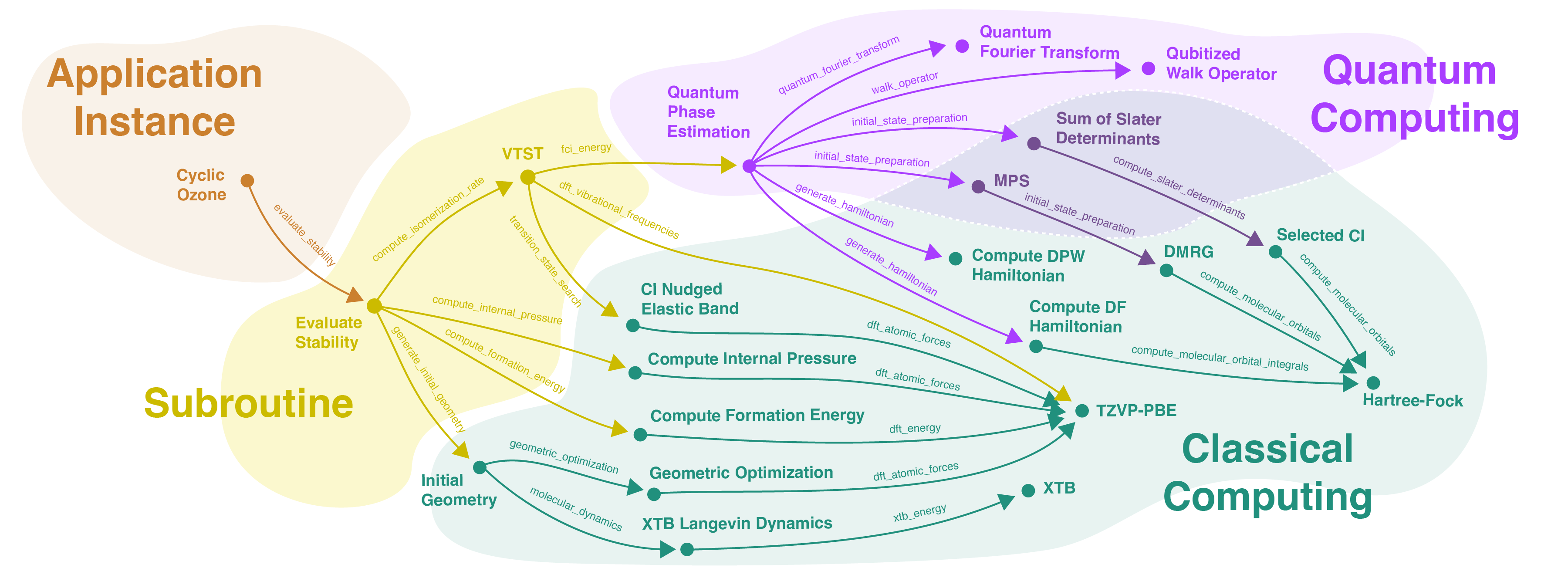}
\end{center}
    \caption{{\bf A Quantum Benchmarking Graph (QBG) for ozone isomerization}.  This callgraph captures a full computational workflow, including both  quantum and classical subroutines. Arrows point to nodes that provide a required action/quantity, as specified  by the edge labels between nodes. Each colored region represents a different level of abstraction, while classical and quantum components are divided into separate  green and purple regions, respectively. Note that the arrows also show how the classical and quantum subroutines exchange data such as molecular geometries, ground state energies, Hamiltonians, and initial classical guesses for the ground state.}
    \label{fig:qbg}
\end{figure}

 We construct \textit{quantum benchmarking graphs} (QBG) to track the cost of our computation.  These are essentially call--graphs at decreasing levels of abstraction, which aggregate the total complexity of all required classical and quantum subroutines (algorithms). The QBG for cyclic ozone is shown in Fig.~\ref{fig:qbg}. The highest level of this hierarchy calls a workflow that implements a solution of our application problem.  While our analysis is based around VTST, this is not the only way to handle the underlying tasks.  That is, other workflows might populate this level of abstraction with alternative computational methods (e.g. algorithms or sub--workflows implementing these algorithms).

\par Working within the VTST workflow, the node \textbf{Evaluate Stability} applies a series of methods that were introduced in \cite{thermal_stability_of_cyclic_ozone, hydrogen_storage, Sharma2025} to oxygen--containing fullerenes. This workflow calls \textbf{VTST}, which requires quantities such as FCI--quality energies and molecular geometries along the reaction pathway. Our implementation uses quantum phase estimation (QPE) to obtain these energies for a few specific geometries. The geometries, in turn, are derived classically by computing a 
minimum energy path between reactants and products using the nudged elastic band (NEB) method \cite{Jonsson1998} and density functional theory (DFT) calculations  (taken specifically as the \textbf{Climbing Image Nudged Elastic Band} (CI-NEB) variant) \cite{Henkelman2000a,Henkelman2000b}.  Note that NEB accepts atomic forces and energies from DFT, which are  passed back as inputs in the QBG.

\par The remaining classical subroutines generate molecular geometries for various endofullerenes as well as a series of stability parameters \cite{hydrogen_storage}.  These include the formation energy $\Delta E$ and an internal pressure $P(\epsilon)$ exerted by encapsulated molecules on the carbon skeleton. The latter is quantified through the bond strain tensor.  Both of these quantities can be obtained using DFT and the geometric arrangement described in \ref{fig:endofullerenes}, where the \textbf{TZVP--PBE} node represents and provides low--level energies and atomic forces. These quantities inform us about the stability of the fullerene cage as it confines the ozone molecules. If the temperature rises too high or the cage is filled with too much oxygen, it may rupture and release the encapsulated ozone molecules \cite{hydrogen_storage}. 

\par Initial geometries are generated by using a sphere packing algorithm to place ozone molecules in a target fullerene.  We can approximate the free volume as a spherical region of radius $R_\text{free} = R_0 - R_\text{vdW}(C)/2$, where $R_0$ is the radius of the cage and $R_\text{vdW}(C)$ is the van der Waals radius of a carbon atom.  A similar strategy is used to choose a radius $r$ for spheres that bound the space surrounding each ozone molecule. The values $R_\text{free}$ and $r$ can then be slightly adjusted to obtain tighter sphere packings. The resulting geometries are first relaxed using a semiempirical method like the extended tight binding approach (XTB; specifically as the GFN1-xTB method) \cite{Grimme2017}.

\par The packed geometries are not realistic, finite--temperature atomic configurations.  To remedy this, the QBG workflow samples coordinates from \emph{ab initio} molecular dynamics (MD) simulations that are based on tight--binding DFT (XTB) and a Langevin thermostat.  We utilize a method based on \cite{Ricci2003, Kuhne2007} as implemented in CP2K \cite{cp2k}. While Langevin dynamics can have unphysical kinetics, they rapidly sample a physically meaningful ensemble.  The inputs for our resource estimates were prepared  using this scheme.  Note that we couple only the fullerene atoms to the Langevin bath and integrate equations of motion using a classical Trotter--Suzuki integrator (coupling frequency $\gamma = 0.005$ fs$^{-1}$; timestep of $\delta = 0.5$ fs) \cite{Ricci2003}.  This gives a natural arrangement where the cage links to the broader environment, with minimal kinetic artifacts for the oxygen atoms \cite{Onuki2019}. This is followed by full DFT relaxation using the PBE density functional and a D3 dispersion correction, alongside the  TZVP basis \cite{Schaefer1994} and GTH pseudopotentials \cite{Goedecker1996}.  A comparable scheme was previously shown to give reasonable geometries for chemically similar systems \cite{3zeta_dft_1, 3zeta_dft_2, entropy_of_ozone}.

\par The last component is a classical algorithm that prepares the initial state for quantum phase estimation. The \textbf{DMRG} and \textbf{Selected CI} methods both accomplish this task and there are efficient methods to load these outputs into quantum computers \cite{selectedci_dmrg, Malz2024}.  However, the  overlap between the initial guess and the true ground state will influence the number of QPE circuit shots (Fig.~\ref{fig: qubitized_phase_estimation}) that are needed to obtain chemical accuracy in our energy estimates (defined as $1.6 \text{ mHa}$).  This critical detail puts fixed--basis electronic structure into the QMA--complete complexity class, which is the quantum analogue of NP--complete.  It is provably hard for quantum computers to solve these problems and that fact can compromise an exponential speedup \cite{speedup_qchem}.  We can always select for classically intractable chemistry problems that are in a ``goldilocks zone'' with sufficient overlap between our guess and the true ground state (e.g., impurity Hamiltonians beyond the free fermion picture and their hidden variants)~\cite{Chapman2020,Elman2021,chapman2023unified}.  A sufficient overlap is likely for the computational tasks that we describe. However, these considerations ultimately mean that QPE is heuristic  and capturing its true performance will require a comparison of different approaches to generating the initial guess.

\subsection{Qubitized Phase Estimation}

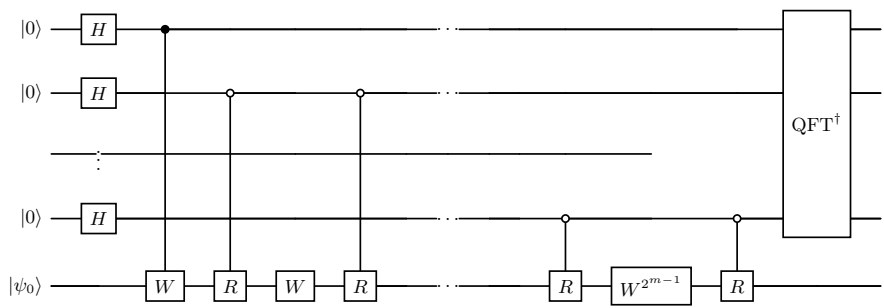
\begin{figure}
\begin{center}
\begin{adjustbox}{scale=.98}  
\begin{quantikz}
\lstick{$\ket{0}$}   & \gate{H} & \ctrl{4} & \qw & \qw & \qw  & \qw&\dots& & \qw &\qw&\qw  &\qw & \gate[4]{\text{QFT}^\dagger} & \qw \\
\lstick{$\ket{0}$}   & \gate{H} & \qw & \octrl{3} & \qw   & \octrl{3}&\qw 
                     &\dots &\qw&\qw &\qw &\qw &\qw &\qw & \qw \\
                     \vdots  \\
\lstick{$\ket{0}$}    & \gate{H}& \qw & \qw & \qw  &\qw &  \qw&\dots& &\qw &\octrl{1} &\qw &\octrl{1}&\qw &\qw  \\
\lstick{$\ket{\psi_{0}}$} \qw & \qw & \gate{W} & \gate{R}  & \gate{W} &\gate{R}&\qw &\dots & &\qw & \gate{R} & \gate{W^{2^{m-1}}} & \gate{R} & \qw&  \qw
\end{quantikz}
\end{adjustbox}
\end{center}    
    \caption{ {\bf Quantum circuit implementing qubitzied phase estimation.} This implementation queries powers of the qubitized walk operator $W$ instead of the standard time evolution oracle. The eigenphases $\phi_k$ of $W$ have a simple functional relation with the eigenenergies $E_k$ of a target Hamiltonian $H$. Specifically, $\phi_k =  \text{arccos}(E_k / \lambda)$, where $\lambda$ is a normalization factor for the block encoding of $H$. This relationship allows us to extract the eigenenergies using phase estimation. }
    \label{fig: qubitized_phase_estimation}
\end{figure}

\begin{figure}
\begin{center}
\begin{adjustbox}{scale=1.05}  
\begin{quantikz}
     \qw & \qw              & \qw       & \qw                  & \octrl{1}                      & \qw                                                        &  \qw & \gate{Z}   &  \qw \\
\lstick{\text{Selection} \;\; }        \qw & \qw \qwbundle{m} & \qw       & \gate{\text{Prep}}   & \gate[wires=2]{\text{Select}}  & \gate{\text{Prep}^{-1}}  &  \qw & \octrl{-1} &  \qw \\
\lstick{Target \;\;\;\; }           \qw & \qw \qwbundle{n} & \qw       & \qw                  & \qw                               & \qw                      &  \qw & \qw        &  \qw
\end{quantikz}
\end{adjustbox}
\end{center}    
    \caption{ {\bf Qubitized Walk Operator.} The walk operator $W$ is implemented via Prepare and Select oracles, which specify a block encoding for the Hamiltonian $H$, and a multicontrolled CZ gate which gives a reflection in each qubitized subspace.  The selection qubits are a set of $m = \lceil \log_2 n \rceil$ ancilla that implement the block encoding and the target is an $n$--qubit state.}
    \label{fig:qubitized_walk_operator}
\end{figure}
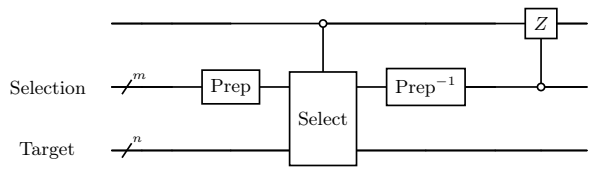

Many ground--state estimation protocols are based on the quantum phase estimation (QPE) algorithm~\cite{kitaev1995quantum}. A qubitized variant of this method can markedly improve runtime efficiency by encoding electronic spectra of the Hamiltonian $H$ in a Szegedy walk operator $W$ \cite{Berry2018, Poulin2018,babbush2018encoding}.  Loosely stated, qubitization describes a strategy that encodes a normalized, non--unitary operator $H/\lambda$ within a designated block of a larger unitary operator $U_H$ \cite{Low2019hamiltonian}.  The factor $\lambda$ is  normalization that ensures unitarity.  This gives an effective two--dimensional Hilbert space for each eigenvalue $E_k$, which is spanned by the eigenstate and its orthogonal complement.  The resulting walk operator assumes a particularly useful form on each space,

\begin{equation}
W = \begin{pmatrix} \frac{E_k}{\lambda} & \sqrt{1 - \left(\frac{E_k}{\lambda}\right)^2} \\
-\sqrt{1 - \left(\frac{E_k}{\lambda}\right)^2} & \frac{E_k}{\lambda}  \end{pmatrix} = e^{\imath \arccos (E_k/\lambda) Y}
\end{equation}

\noindent where $Y$ denotes the corresponding Pauli operator.  Notably, this gives a simple functional relationship $\phi_k =  \text{arccos}(E_k / \lambda)$ between eigenphases and eigenenergies for a given Hamiltonian.  By extending this convention to QPE, an eigenenergy can be estimated to a precision of $\Delta E$ by using $m = \lceil \log_2 (\sqrt{2}\pi\lambda / 2\Delta E)\rceil$ precision qubits (ancilla). A circuit implementing this QPE approach is depicted in Fig.~\ref{fig: qubitized_phase_estimation} and a circuit for the qubitized walk  $W$ is given in Fig.~\ref{fig:qubitized_walk_operator}

 Here, the data register is prepared in a state that overlaps significantly with the ground state. The success probability in postselection depends on this overlap, so it is important that it is high (see \cite{Gharibian2022} for an analysis of the resulting complexity). The walk operator is assembled using \texttt{PREPARE} and \texttt{SELECT} oracles, which is a standard strategy to block encode Hamiltonians following the Linear Combination of Unitaries (LCU) access model~\cite{Childs2012, Berry2015, Low2019hamiltonian}.   At the end of the circuit, an inverse quantum Fourier transform (QFT) allows the accumulated phase to read off the ancilla in a bitwise manner. We defer to the discussion in  ~\cite{babbush2018encoding} for further details.

\subsection{Hamiltonian Encodings}
Endofullerenes such as (12 O$_3$)@C$_{180}$ are larger than the molecules typically studied using post--Hartree--Fock techniques.  These molecules also lack the clear periodic structure of condensed matter systems. These factors present a challenge for algorithm selection. We describe and estimate resources using multiple methods, allowing us to capture the  landscape of possible Hamiltonian encodings and the resulting effect overhead. All of these methods could provide sufficient accuracy if the underlying scale parameters (e.g., size of the active space, number of dual plane waves) are large enough to capture electronic wavefunction. 

\subsubsection{Local Basis: Double--Factorized (DF) Encoding}
Small--molecule quantum chemistry is typically handled using a local orbital basis.  While a full set of these classical orbitals could be included in our quantum calculation, this generally gives excessive overhead and a diminishing return.  We will instead use an \emph{active space} approach,  which conjoins an effective Hamiltonian with a smaller set of problem--adapted  orbitals (often tens to hundreds for small molecules). We define an initial, classical electronic structure using Hartree--Fock with a minimal STO--3G~\cite{hehre1969self} basis set for fullerene carbons and a larger cc--pVDZ~\cite{dunning1989gaussian} basis for the oxygen atoms. The result describes (12 O$_3$)@C$_{180}$ using a total of 1404 orbitals.  This output is then used to define active spaces of increasing size via the AVAS method \cite{AVAS}, as facilitated by the PySCF~\cite{sun2018pyscf} code.  Our procedure delivers coefficients for the standard quantum chemistry Hamiltonian,

\begin{equation}\label{eq:ham_qchem}
    H = \sum_{ij,\sigma} h_{ij} a^\dagger_{i\sigma} a _{i\sigma} 
 + \frac{1}{2}\sum_{ijkl,\sigma\rho} h_{ijkl} a^\dagger_{i\sigma} a^\dagger_{j\rho}a_{k\rho} a_{l\sigma},
\end{equation}
where $h_{ij}$ and $h_{ijkl}$ are the one-- and two--electron integrals, $\sigma$ and $\rho$ index spin, and $a_{p\sigma}$ are fermionic creation and annihilation operators. There is considerable overhead when encoding this Hamiltonian for use on a quantum computer.  Instead, we opt to use a more resource--efficient encoding known as double--factorization~\cite{von2021quantum}.

To understand this, note that the fourth--order Coulomb tensor $ h_{ijkl} $ can be rearranged into a $ N^2/4 \times N^2/4 $ matrix, where $N$ is the number of \textit{spin orbitals}.  This is termed the electronic repulsion integral (ERI) matrix $A$, which is positive semi-definite and exhibits a rank $ L = O(N) $ for chemical systems. Diagonalizing $A$ leads to a matrix decomposition in terms of an auxiliary tensor $\mathcal{L}$ such that~\cite{Motta2021}:
\begin{eqnarray}
A = \sum_{\ell=0}^{L-1} (\mathcal{L}^{(\ell)})^2 = \sum_{\ell=0}^{L-1} \sum_{ijkl=0}^{N/2-1} \mathcal{L}_{ik}^{(\ell)} \mathcal{L}_{jl}^{(\ell)} a^\dagger_{i} a_{k}a^\dagger_{j} a_{l} .
\end{eqnarray}
Each matrix $\mathcal{L}^{(\ell)}$ can then be diagonalized, giving a set of eigenvalues $\{\lambda^{(\ell)}_m \}$ and diagonalizing unitary $U^{(\ell)} $ so that the double factorized Hamiltonian $H_{\text{DF}}$ is constructed as:

\begin{eqnarray}
    \label{eq:df-ham}
H_{\text{DF}} &=& \sum_{ij,\sigma} \tilde{h}_{ij} a^\dagger_{i\sigma} a_{j\sigma}+\frac{1}{2} \sum_{\ell=0}^{L-1} \left(\sum_{ij, \sigma} \sum_m \lambda_m^{(\ell)}  U_{m,i}^{(\ell)} U_{m,j}^{(\ell)} a^\dagger_{i\sigma} a_{j\sigma}\right)^2, \nonumber\\
\end{eqnarray}
where $\tilde{h}_{ij} \equiv h_{ij} - \frac{1}{2}\sum_{l}h_{illj}$ comes from the reordering of the creation and annihilation operators. An advantage of this representation is that the sum over $m$ can be truncated such that a low--rank approximation is obtained, further reducing the number of terms needed to describe the Hamiltonian. This double factorized Hamiltonian $H_{\text{DF}}$ can be efficiently translated into the walk operator $W$ that is required for QPE, as described in Ref.~\cite{von2021quantum}.

\subsubsection{Semilocal Basis: Dual Plane--Wave Encoding}

\begin{figure}[t]
    \centering
    \includegraphics[scale=0.3]{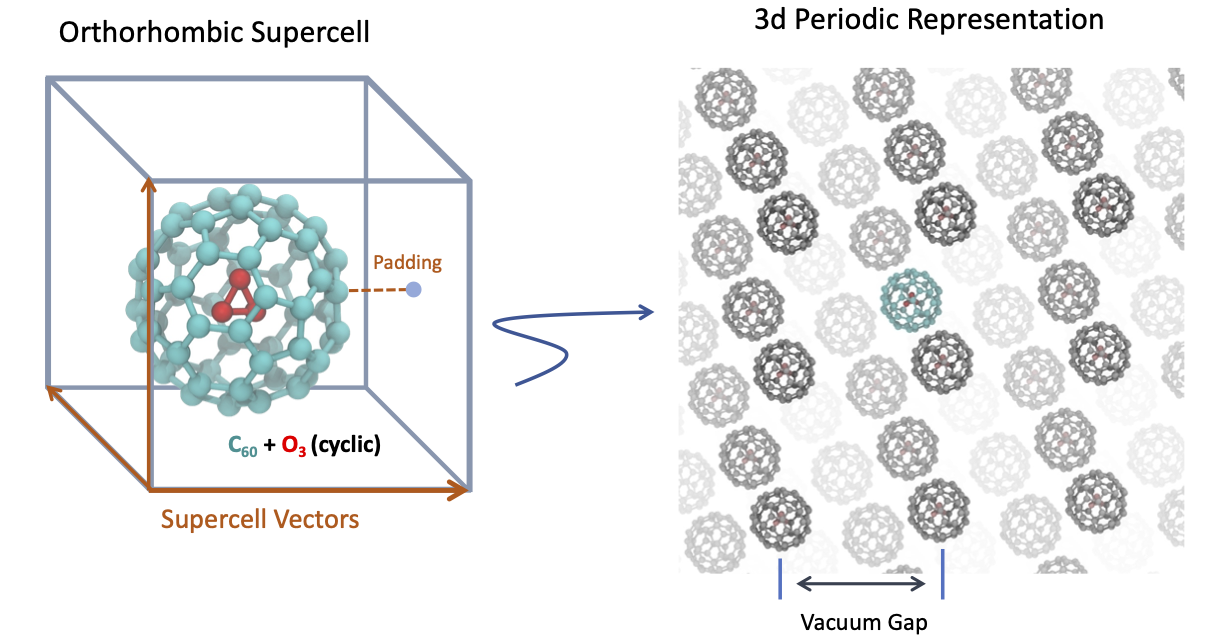}
    \caption{{\bf Endofullerene in a periodic supercell.} Geometry of an ozone--containing endofullerene $O_3$@C$_{60}$ embedded in an orthorhombic supercell. Note the vacuum padding that flanks the fullerene cage.  This ultimately defines the geometric separation between periodic images. }
    \label{fig:O_3@C60_in_box}
\end{figure}

\par The  dual plane--wave (DPW) basis is constructed as the Fourier transform of a nonlocal plane--wave (PW) basis, and corresponds to semilocal functions that are arranged on a regular real--space grid. This scheme has certain advantages for quantum algorithms --- such as a favorable  $O(N^2)$ scaling for terms in the Coulomb tensor ---  though it assumes a periodic system.  We can nonetheless treat isolated molecules like our endofullerenes by separating periodic images with a vacuum gap.  This gap should be large enough to attenuate electrostatic self--interaction and to prevent wavefunction `leakage' across periodic boundaries (Fig.~\ref{fig:O_3@C60_in_box}).  The $N$--spin--orbital (basis element) DPW electronic structure Hamiltonian takes the form:

\begin{align*}
    \label{eq: DPW_2ndQuantize_ElectronicHam}
    H &= \sum_{pq,\sigma} T(p-q) a^\dagger_{p\sigma}a_{q\sigma} + \sum_{p\sigma} U(p) n_{p\sigma} + \sum_{(p\alpha)\neq(q\beta)} V(p-q)n_{p\alpha}n_{q\beta} \\
    &=  \underbrace{\frac{1}{N} \sum_{pq,\sigma, \nu} k_\nu^2 \, \cos \big[ k_\nu \cdot r_{p-q} \big] a^\dagger_{p\sigma}a_{q\sigma}}_{\text{kinetic}} 
    -  \underbrace{\frac{4\pi}{\Omega} \sum_{p\sigma, j\nu \neq 0 } \frac{\zeta_j \, \cos \big[ k_\nu \cdot (R_j - r_p)\big]}{k_\nu^2} n_{p\sigma}}_{\text{electron-ion}} \\
    &+\underbrace{ \frac{2\pi}{\Omega} \sum_{(p\alpha)\neq(q\beta)} \frac{\cos \big[k_\nu \cdot r_{p-q} \big]}{k_\nu^2} n_{p\alpha}n_{q\beta}}_{\text{electron-electron}}
\end{align*}

\noindent where the fermionic operators $a^\dagger_{p\sigma}$ and $a_{p\sigma}$ now act on a spatial spin orbital $p$ and where ${n_{p\sigma} = a^\dagger_{p\sigma}a_{p\sigma}}$ is the number operator. Each orbital is associated with an orbital centroid,
\[
    r_p = p\left(\frac{2\Omega}{N}\right)^{1/3},
\]
where $\Omega$ is the computational cell volume and $R_j$ is the position of atom $j$ with atomic number $\zeta_j$. The momentum modes are defined as,
\[
k_\nu = 2\pi\frac{\nu}{\Omega^{1/3}},
\]
with $\nu \in [-(N/2)^{1/3}, (N/2)^{1/3}]^{\times 3}$. This closely follows the notation of \cite{babbush2018encoding}. As noted earlier, the number of electron--electron interaction terms scales as $O(N^2)$, while they scale as $O(N^3)$ in a plane--wave basis and $O(N^4)$ using Gaussian type orbitals. While kinetic terms also scale as $O(N^2)$ in the DPW basis (as opposed to $O(N)$ as in the plane--wave basis) this increase is offset by reduced complexity of the Coulomb tensor. A key parameter for this method is the distance $a_0$ between adjacent points in the real--space basis grid.  In principle, this lengthscale can be loosely related $a_0 \approx \sqrt{2\pi^2/E_\text{cut}}$  to the energy cutoff $E_\text{cut}$ for some equivalent plane--wave calculation (i.e., formally via a Fourier transform).  However, there will be some degree of  ambiguity when translating convergence with momentum--space pseudopotentials to their real--space counterparts.

\section{Resource Estimates}
\label{sec: Resource Estimates}

\par Asymptotic scaling can give a misleading impression of computational overhead, since many problems are of a size where algorithmic prefactors and subleading contributions can be relevant.  This becomes especially pronounced when we consider specific Hamiltonian encodings.  We avoid this complication by explicitly counting resources using a detailed profiling strategy and explicit circuit constructions.  These tasks are facilitated by the Azure Quantum Resource Estimator~\cite{Azure_Quantum_Resource_Estimator,beverland2022assessing}, PennyLane's resource estimation framework \cite{bergholm2018pennylane} and the pyLIQTR toolkit \cite{pyliqtr}, respectively.

\subsection{Tools and methodology}
\subsubsection{AzureQRE}

The Azure Quantum Resource Estimator (AzureQRE)~\cite{beverland2022assessing,Azure_Quantum_Resource_Estimator} is a component of the Azure Quantum Development Kit~\cite{Microsoft_Azure_Quantum_Development}, which handles both logical and hardware--level estimation. At the application level, we set an accuracy threshold of 1 mHa for energy determinations. This is below the standard chemical accuracy  of 1.6 mHa.  The hardware level estimates use the  \texttt{qubit\_gate\_ns\_e4} model within AzureQRE, which assumes an optimistic superconducting  device (e.g., transmon qubits) having 50 ns gate times,  100 ns measurement times, and  error rates of $10^{-4}$ for both  Clifford and non--Clifford mechanisms. We set a total error budget of  1\%. Logical circuits implementing the double--factorized qubitization algorithm~\cite{von2021quantum} are complied within Q\# and passed into the physical resource estimator. Fault--tolerance is handled by through a surface code~\cite{fowler2012surface} implementation  on a 2D nearest--neighbor layout that is capable of parallel operations. The overhead of logical qubit movement and multi--qubit measurements using lattice surgery operations is included \cite{Horsman2012, Litinski2019}. $T$--gates are implemented through a magic state distillation protocol in dedicated factories~\cite{bravyi2005universal}.  More detail regarding the architectural assumptions used in AzureQRE are described elsewhere~\cite{beverland2022assessing}.

The truncation procedure of the double factorized Hamiltonian in Eq.~(\ref{eq:df-ham}) follows the incoherent scheme described in \cite{von2021quantum}, which assumes that the errors of truncated terms add incoherently by a sum--of--squares. Specifically, given the difference between the low--rank approximated Hamiltonian $\tilde{H}_{\text{DF}}$ and the numerically exact $H_{\text{DF}}$ can be bounded in the spectral norm as,
\begin{equation}
\lVert H_{\text{DF}} - \tilde{H}_{\text{DF}} \rVert \leq \lVert \mathcal{L}^{(\ell)} \rVert_{SC} | \lambda_m^{(\ell)} |,
\end{equation}
where $\lVert \mathcal{L}^{(\ell)} \rVert_{SC}$ indicates the Schatten norm. Truncation is carried out such that,
\begin{equation}
\sqrt{\sum_{\ell,m}(\lVert \mathcal{L}^{(\ell)} \rVert_{SC} | \lambda_m^{(\ell)} |)^2} \leq \epsilon_{\text{trunc}},
\end{equation}
where $\epsilon_{\text{trunc}}$ is the target low--rank approximation error.

\subsubsection{PennyLane Resource Estimator}
\par We also generated logical--level resource estimates using the PennyLane framework~\cite{bergholm2018pennylane}.
Our workflow follows the end--to--end procedure documented in the PennyLane resource--estimation tutorial~\cite{JaySoni2026}, which standardizes the choices of Hamiltonian representation, truncation thresholds, and QPE error targets used throughout this work. Specifically, we used the \texttt{qml.estimator} module, which implements analytical costing formulas for quantum phase estimation (QPE) tailored to Hamiltonian representations common in quantum chemistry. For second--quantized molecular Hamiltonians, PennyLane provides the \texttt{qml.estimator.DoubleFactorization} resource class, which estimates the required number of logical qubits and non--Clifford gates (reported as Toffoli--type costs in the underlying implementation) to perform QPE using a double--factorized Hamiltonian representation.

PennyLane's \texttt{qml.estimator}  module uses a compact representation for operators in order to estimate the costs of algorithms without building the associated complete molecular Hamiltonians. This allows for quick feedback for investigating logical algorithmic costs, typically on the order of milliseconds independent of the system size. The estimator takes as input the one-- and two--electron integrals in a molecular orbital basis. Internally, the two--electron tensor is expressed in chemist notation and then decomposed via the two--stage (double) factorization procedure described in Ref.~\cite{von2021quantum}. In the first stage, the four--index two--electron tensor is approximated as a sum of rank--$R$ symmetric factors,
\begin{equation}
V_{ijkl} \;\approx\; \sum_{r=1}^{R} L^{(r)}_{ij}\, L^{(r)T}_{kl},
\end{equation}
followed by a second--stage diagonalization of each $L^{(r)}$ that yields an average retained eigenrank $M$ after truncation. These effective ranks $(R,M)$, together with the Hamiltonian 1--norm $\lambda$ computed from the retained factors, determine the asymptotic cost of the qubitization/QPE procedure through the closed--form expressions implemented in the estimator.

\par In all PennyLane DF estimates reported here, we fixed the truncation thresholds for \emph{both} stages of the factorization to $10^{-5}$:
(i) the factor (first--stage) truncation threshold \texttt{tol\_factor} and
(ii) the eigenvalue (second--stage) truncation threshold \texttt{tol\_eigval},
\begin{equation}
\texttt{tol\_factor} \;=\; 10^{-5},
\qquad
\texttt{tol\_eigval} \;=\; 10^{-5}.
\end{equation}
Operationally, \texttt{tol\_factor} discards negligible first--stage factors, while \texttt{tol\_eigval} discards small eigenvalues (and their eigenvectors) in the diagonalization of each retained factor, thereby controlling $(R,M)$ and the resulting $\lambda$.

\par To match the application--level accuracy used throughout this work, we set the target QPE energy error parameter in PennyLane to 1~mHa (i.e., \texttt{error}$=10^{-3}$~Ha). The outputs we report from PennyLane are the resulting logical qubit counts and non--Clifford gate counts implied by the DF costing formulas.

\subsubsection{pyLIQTR}

\par  The pyLIQTR toolkit implements a range of quantum algorithms that can be analyzed at the logical level.  A distinguishing characteristic of pyLIQTR is its hierarchical nature, where the top--level circuit is an entire qubitized algorithm.  This be  decomposed in stages, yielding progressive subcircuits that are  ultimately translated into the Clifford+$T$ gate set.  Such modularity leads to efficient resource estimation, since repeated elements can be analyzed once and cached for later use. We use pyLIQTR for our DF and DPW resource estimates.

\par The DPW grid spacing was chosen to match energy cutoffs from a previously benchmarked PW pseudopotential scheme. Ultrasoft pseudopotential calculations suggest that an $E_\text{cut} = 40$ Ry cutoff would be sufficient for this system, which corresponds to a separation of $a_0 = 0.526$ \,\AA\, between the loci for DPW basis functions.   We assume that an equivalent DPW pseudopotential can be absorbed into the Hamiltonian coefficients, but do not attempt to make this quantitative (this should only impact the encoding normalization).   Molecules were isolated from their periodic images using a 10\,\AA\, vacuum padding and explicit Hamiltonian coefficients were generated from  target molecular geometries using pyLIQTR's PEST module.  These data were passed to the pyLIQTR circuit generation framework, which implements a block encoding for DPW electronic structure with linear $T$--complexity in problem size~\cite{babbush2018encoding}.  The result defines a walk operator for qubitized phase estimation.  Explicit circuits were   generated assuming a target accuracy of 1 mHa for energy estimations.  This formed the foundation for our resource estimates.

\par  For a given circuit, the resources reported by pyLIQTR include logical qubits, $T$--gates, and Clifford gates. This includes contributions from rotation synthesis, which are estimated using a heuristic model that depends on the user specified rotation gate precision. Our targeted energy error of $10^{-3}$ implies that an upper bound on this precision should be on the order of $10^{-6}$.  We ultimately adopt a precision of $10^{-10}$, which ensures that rotation synthesis will be a limiting factor.  However, this also implies that our resource estimate values will constitute an upper bound.

\subsection{Resource Estimation Results}

\begin{figure}
    \centering
    \includegraphics[width=\linewidth]{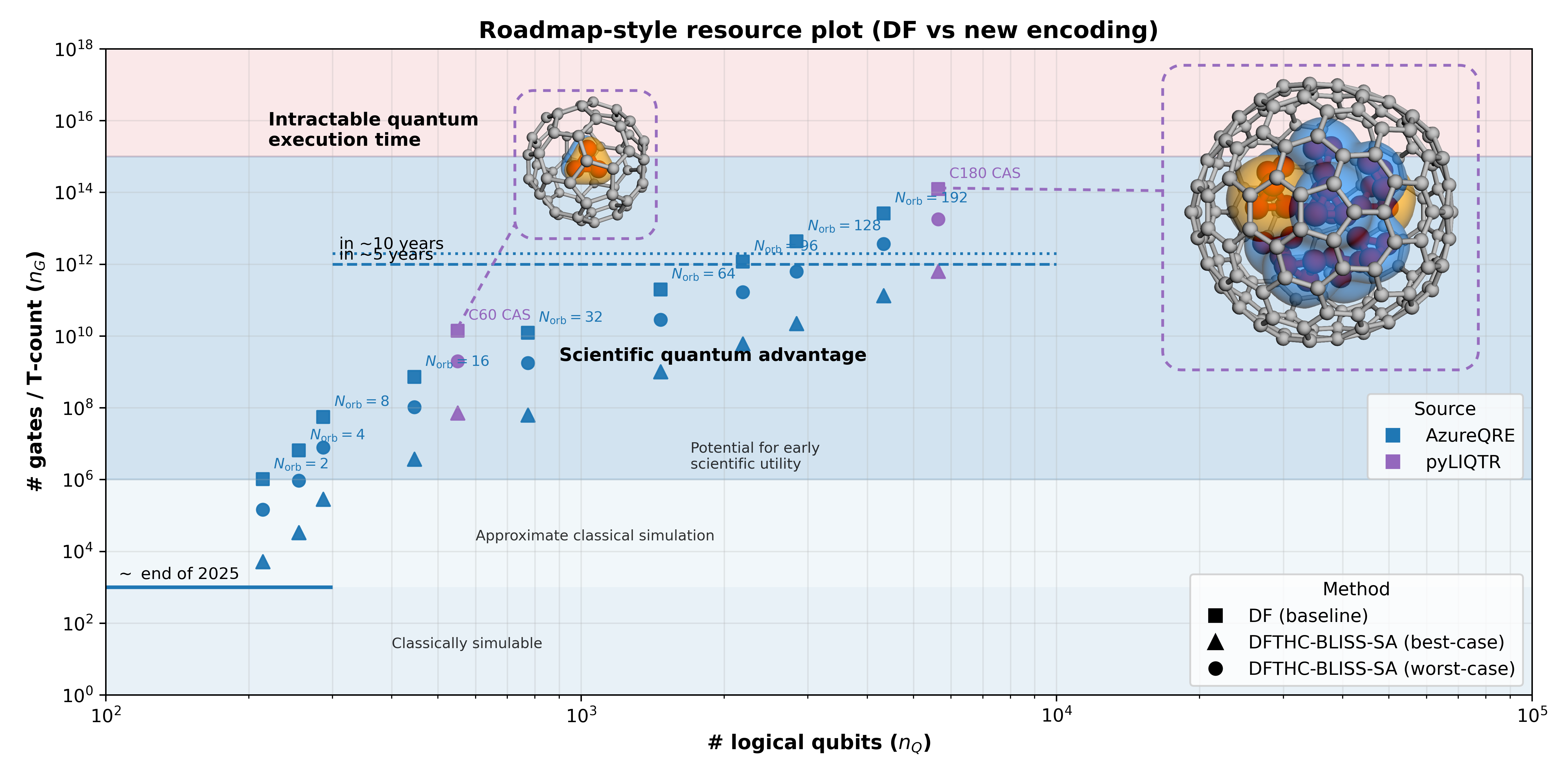}
    \caption{Double factorized 2nd baseline $T$-gate estimates ($n_G$) versus logical qubits ($n_Q$) with boundaries for classical tractability, quantum advantage and intractable quantum runtime as defined in \cite{camps2025quantumcomputingtechnologyroadmaps}. In addition to resource estimates, we show the range of possible savings that can be realized by using the state-of-the-art chemical Hamiltonian encoding ``DFTHC-BLISS-SA'' \cite{Low_2025}. The target low-rank approximation error in the AzureQRE was taken to be $\epsilon_\mathrm{trunc}=0.001$. The single and double factorization tolerances for the pyLIQTR resource estimates were taken to be $\epsilon_\text{SF}=\epsilon_\text{DF}=10^{-5}$.}
    \label{fig:joe_plot}
\end{figure}

\subsubsection{Local Basis: Double--Factorized Hamiltonian}

\par The logical quantum resource estimates for a single--shot QPE calculation are depicted in Fig.~\ref{fig:joe_plot}.  These data assume active spaces of varying size for the (12 O$_3)$@C$_{180}$ system and a double--factorized Hamiltonian encoding. At the logical level, we see that this  algorithm has a spatial overhead of roughly 20 logical qubits per orbital.  The temporal resource requirements are also substantial, with the smallest active space --- which is classically trivial --- requiring approximately $10^6$ $T$--gates.  Conversely, the largest, maximal--performance active space requires over 4000 logical qubits and $2.6 \times 10^{13}$ $T$--gates.  This leads to hefty physical overhead from the error--correction layer, and we estimate that approximately $6 \times 10^6$ physical qubits for a $d = 19$ surface code implementation.

\subsubsection{Electrostatic Embedding Techniques}

\par Our energy estimates correspond to a fixed atomic geometry. Moreover, the carbon-based skeleton is expected to contribute only weakly to the \emph{correlated} electronic structure of the encapsulated ozone species: to leading order, the cage acts as a rigid nanocontainer that shifts orbital energies and hybridization patterns through electrostatic confinement and polarization, rather than by participating directly in bond rearrangement. This motivates an electrostatic-embedding perspective in which the fullerene environment is treated at a lower level (e.g., as a fixed set of point charges or a frozen mean-field potential), while the quantum calculation focuses on the oxygen subspace that controls isomerization and reactivity.

\par Within this embedding viewpoint, we construct oxygen-only active spaces using the atomic valence active space (AVAS) procedure~\cite{AVAS} as implemented in PySCF~\cite{sun2018pyscf}. Concretely, starting from a reference mean-field calculation (here, a mixed-basis setup with a minimal description on the carbon scaffold and a larger basis on oxygen), AVAS selects molecular orbitals by maximizing their overlap with a chosen set of target atomic valence orbitals. We target the O $2s$ and $2p$ valence manifold, so that the retained active orbitals are those most “oxygen-like” in the sense of AVAS projection. Operationally, we (i) specify the oxygen $2s/2p$ valence AOs as the target space, (ii) run AVAS on the resulting SCF orbitals, and (iii) retain the resulting active orbitals up to the point where the AVAS overlap metric indicates negligible additional oxygen character. This produces the oxygen-only active spaces reported in Table~\ref{tab:df_vs_new_gatecounts}, including CAS(12o,18e) for $\mathrm{O}_3@C_{60}$ and CAS(132o,198e) for $(12\,\mathrm{O}_3)@C_{180}$, which are physically motivated by focusing correlation on the oxygen valence degrees of freedom while treating the fullerene primarily as an embedding environment.

\par The practical advantage is that this construction sharply reduces quantum resources by shrinking the number of (spin-)orbitals that enter the correlated Hamiltonian, which in turn lowers both the logical qubit footprint and the gate complexity of qubitized QPE for a fixed target energy precision. These embedding-driven reductions are complementary to algorithmic progress in Hamiltonian encodings: for a \emph{fixed} chemically motivated active space (such as oxygen-only AVAS), state-of-the-art encodings can further push gate counts down by substantial constant factors, tightening the overall feasibility window without changing the underlying chemical model \cite{Low_2025}.

\par As an additional consistency check, we report DF resource estimates from both PennyLane and pyLIQTR, which agree at the level of qualitative scaling and order-of-magnitude trends across the active-space ladder. PennyLane’s estimator provides rapid, closed-form logical costing for qubitized QPE under a specified double-factorization truncation policy, making it well suited for sweeping active-space choices and error targets. In contrast, pyLIQTR explicitly \emph{constructs} the corresponding algorithmic circuits (including the qubitization primitives and their subroutines), compiles them down to a fault-tolerant gate set, and then counts resources from the compiled representation. Because pyLIQTR resolves implementation details that are necessarily abstracted in analytic models—such as concrete oracle constructions, arithmetic choices, and rotation synthesis overhead—it can shift absolute gate counts relative to PennyLane while preserving the same overall resource trajectory. The resulting side-by-side comparison therefore serves two roles in our workflow: PennyLane provides a fast ``first-pass’’ estimate for exploring design space, while pyLIQTR validates and refines these estimates by exposing how circuit-level design decisions translate into observable changes in $n_Q$ and $n_G$ in Table~\ref{tab:df_vs_new_gatecounts}.

\begin{table}[t]
\renewcommand{\arraystretch}{1.12}
\setlength{\tabcolsep}{5pt}
\begin{tabular}{llrrr}
\hline
System & Estimator & $n_Q$ (logical) & $n_G^{\mathrm{DF}}$ (T gates) & $n_G^{\mathrm{DFTHC+BLISS+SA}}$ range \\
\hline
$(12\,\mathrm{O}_3)@C_{180}$, $N_\mathrm{orb}=2$   & AzureQRE & 214  & $1\times 10^{6}$   & $[5.2\times 10^{3},\,1.5\times 10^{5}]$ \\
$(12\,\mathrm{O}_3)@C_{180}$, $N_\mathrm{orb}=4$   & AzureQRE & 255  & $6.5\times 10^{6}$ & $[3.4\times 10^{4},\,9.4\times 10^{5}]$ \\
$(12\,\mathrm{O}_3)@C_{180}$, $N_\mathrm{orb}=8$   & AzureQRE & 287  & $5.5\times 10^{7}$ & $[2.8\times 10^{5},\,7.9\times 10^{6}]$ \\
$(12\,\mathrm{O}_3)@C_{180}$, $N_\mathrm{orb}=16$  & AzureQRE & 446  & $7.3\times 10^{8}$ & $[3.7\times 10^{6},\,1\times 10^{8}]$ \\
$(12\,\mathrm{O}_3)@C_{180}$, $N_\mathrm{orb}=32$  & AzureQRE & 773  & $1.2\times 10^{10}$& $[6.3\times 10^{7},\,1.8\times 10^{9}]$ \\
$(12\,\mathrm{O}_3)@C_{180}$, $N_\mathrm{orb}=64$  & AzureQRE & 1470 & $2\times 10^{11}$  & $[1\times 10^{9},\,2.8\times 10^{10}]$ \\
$(12\,\mathrm{O}_3)@C_{180}$, $N_\mathrm{orb}=96$  & AzureQRE & 2190 & $1.2\times 10^{12}$& $[6\times 10^{9},\,1.7\times 10^{11}]$ \\
$(12\,\mathrm{O}_3)@C_{180}$, $N_\mathrm{orb}=128$ & AzureQRE & 2840 & $4.4\times 10^{12}$& $[2.2\times 10^{10},\,6.2\times 10^{11}]$ \\
$(12\,\mathrm{O}_3)@C_{180}$, $N_\mathrm{orb}=192$ & AzureQRE & 4330 & $2.6\times 10^{13}$& $[1.3\times 10^{11},\,3.7\times 10^{12}]$ \\
$\mathrm{O}_3@C_{60}$, CAS(12o, 18e)               & PennyLane & 588  & $1.3\times 10^{9}$ & $[6.7\times 10^{6},\,1.9\times 10^{8}]$ \\
$\mathrm{O}_3@C_{60}$, CAS(12o, 18e)               & pyLIQTR & 550  & $1.4\times 10^{10}$& $[7.1\times 10^{7},\,2\times 10^{9}]$ \\
$(12\,\mathrm{O}_3)@C_{180}$, CAS(132o, 198e)      & PennyLane & 5488 & $2.4\times 10^{13}$& $[1.2\times 10^{11},\,3.4\times 10^{12}]$ \\
$(12\,\mathrm{O}_3)@C_{180}$, CAS(132o, 198e)      & pyLIQTR & 5642 & $1.2\times 10^{14}$& $[6.4\times 10^{11},\,1.8\times 10^{13}]$ \\
\hline
\end{tabular}
\caption{DF baseline $T$-gate estimates ($n_G$) versus logical qubits ($n_Q$), and the implied new-encoding gate-count range
$n_G^{\mathrm{DFTHC+BLISS+SA}} \in [n_G/f_{\max},\, n_G/f_{\min}]$
using improvement factors $f \in [7,195.2]$ (interpreted as $f=n_G/n_G^{\mathrm{DFTHC+BLISS+SA}}$) when using state-of-the-art chemical Hamiltonian encodings \cite{Low_2025}. The target low-rank approximation error in the AzureQRE was taken to be $\epsilon_\mathrm{trunc}=0.001$. The single and double factorization tolerances for PennyLane and pyLIQTR resource estimates were taken to be $\epsilon_\text{SF}=\epsilon_\text{DF}=10^{-5}$.}
\label{tab:df_vs_new_gatecounts}
\end{table}

\subsubsection{Semilocal Basis: Dual Plane--Wave Hamiltonian}

\par
\begin{figure}
    \centering
    \label{fig: t_count_dpw}
    \includegraphics[width=\textwidth]{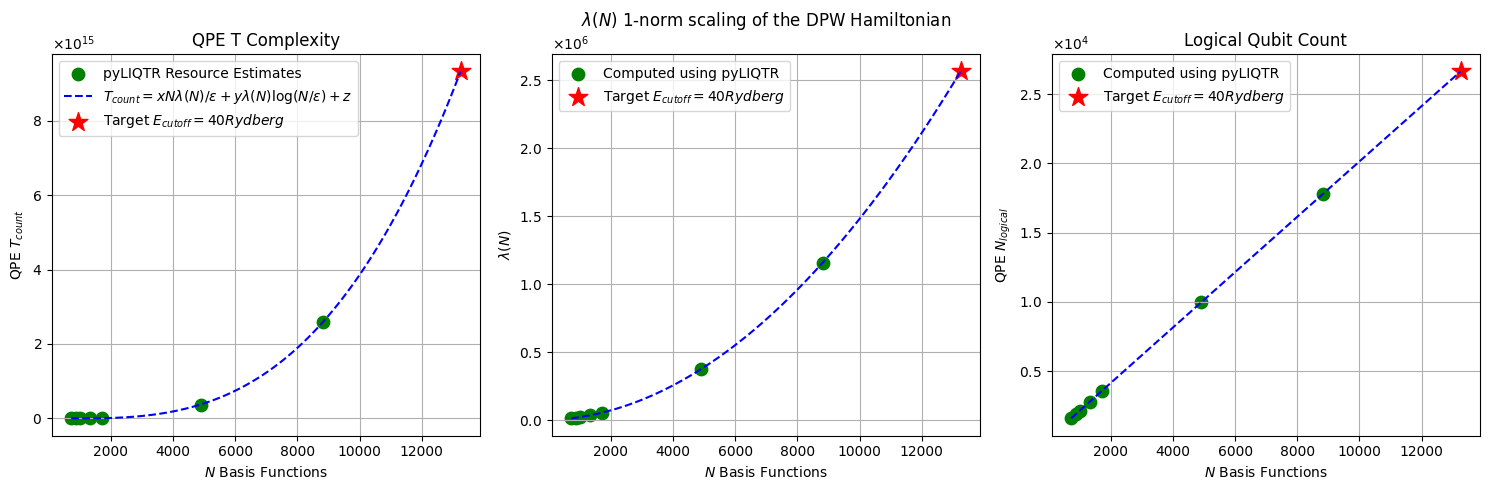}
    \caption{ {\bf Resource estimates for QPE in a dual plane--wave basis.}. Estimates are provided for the O$_3(D{3h}$)@C$_{60}$ system, using an encoding with a linear $T$--complexity in the number of basis functions. Data capture scaling of (a) temporal overhead via the $T$--count; (b) the normalization factor $\lambda$ for the block encoding; and (c) the number of logical qubits that are required for a single--shot calculation.   }
    \label{fig:DPW_resource_estimates}
\end{figure}

 The overhead for QPE with a nonlocal DPW basis is presented in Fig.~\ref{fig:DPW_resource_estimates} and Table~\ref{tab:df_vs_new_gatecounts}.  These estimates are derived using explicit circuits from pyLIQTR, including a block encoding with linear $T$--complexity in orbital count.  Here, the number of basis functions is determined by the DPW grid spacing, which we map to an equivalent plane--wave cutoff.   The overhead for these calculations would be substantially larger than those based on the local basis scheme.  The increased cost is associated with the number of basis functions required for the real--space grid, which includes a large number of points in the vacuum layer.  This is invariably less efficient than the effective models based on active spaces from the preceding sections.  We estimate that for the large fullerene system $(12 \text{ O}_3)@C_{180}$ modelled using an oxygen-only complete active space (CAS) of $N_o=132$ spatial orbitals and $N_e=198$ electrons, we can achieve $T$--gate counts between $10^{11}$ and $10^{13}$ in order to estimate the ground state of the system, with a spatial overhead of $5642$ logical qubits using modern electronic structure quantum circuit encodings \cite{Low_2025}.

\section{Conclusion}

\par This contribution captures an industrially--relevant application for quantum computation, namely assessing the stability of cyclic ozone in an endofullerene matrix.  This is flanked by a classical workflow that addresses geometry of oxygen--containing endofullerenes and prepares inputs for its quantum counterpart.  While our estimates have only explicitly addressed the cost of calculations for isomerization, the single--point energy evaluations used to describe photoisomerization on an excited state surface would have comparable overhead.

\par The  utility of this workflow would be maximized as a high--accuracy ``black--box'' for electronic structure predictions. This could support the direct, quantitative design of new fuel mixtures through a pseudo--combinatorial search over fullerene geometries and atomic configurations.  However, this would also require a large number of calculations with nontrivial resource overhead.  We argue that a less ambitious approach would have comparable utility.  Here, a curated set of calculations would explore thermal stability or photoactivation and derive heuristics that guide experiment.  The same principle would extend to other reactions, such as those underlying fullerene surgery~\cite{synthesis_of_endofullerenes} and the reaction of oxygen equivalents with the fullerene environment.  This follows the general spirit of our workflow, where techniques like VTST are used to simplify rate predictions and leverage classical methods where they have sufficient accuracy.  That is, we are able to maximize efficiency by limiting quantum computation to cases where it would be truly impactful.

\par Endofullerenes present an interesting problem scale for quantum computation.  These are spatially isolated systems with computational tasks that involve the reactivity of small molecules. However, the number of atoms is much larger than many quantum--relevant problems in small--molecule chemistry, such as catalysis.  We captured the overhead for these systems using several algorithmic techniques --- including encodings for local and  semilocal bases --- as well as an electrostatic embedding strategy.  Our classical estimates are in line with other efforts, which have found that significant quantum resources will be required to capture reactivity~\cite{otten2023qrechem,Reiher_femoco,goings2022reliably}.  That being said, there remains a reason for optimism.  A  calculation with entry--level utility (e.g., a 64--orbital active space) may require roughly $10^{11}$ $T$--gates per shot with a predicted runtime of 13.43 days.  This lies just beyond the scale commonly accessed using classical tensor--network methods (these are generally approximate).  While the largest active space (192 orbitals) would require on the order of $10^{13}$ $T$--gates and a runtime of roughly three years, it is not unreasonable that algorithmic improvements could ultimately reduce this by two orders of magnitude. Recent developments based on tensor hypercontraction \cite{first_quantized_pw,hypercontraction} and compressed double--factorization \cite{Cohn2021} have yielded more efficient Hamiltonian encodings, and alternative methods invoking first--quantized electronic structure \cite{first_quantized_pw} can lower logical overhead.  To put this in context, tensor hypercontraction only required $\sim1500$ logical qubits and $7 \times 10^9$ Toffoli gates  \cite{goings2022reliably} to describe reactivity at the catalytic center of cytochrome p450 enzymes.  Given the assumptions underlying our analysis, the associated single--shot runtime would be comparable to relatively trivial calculations in classical quantum chemistry.

\section*{Acknowledgements}
All authors except JEE, KJM, and KMO acknowledge support by the Defense Advanced Research Projects Agency under Contract No. HR001122C0074.  JEE, KJM, and KMO separately acknowledge support from the Defense Advanced Research Projects Agency under Air Force Contract No. FA8702-15-D-0001. Any opinions, findings and conclusions or recommendations expressed in this material are those of the authors and do not necessarily reflect the views of the Defense Advanced Research Projects Agency. MJB acknowledges the support of the ARC Centre of Excellence for Quantum Computation and Communication Technology (CQC2T), project number CE17010001.

The authors thank John Carpenter for support in creating high-resolution figures for this work and Marc Dvorak and Steven Cotton for helpful discussions. The authors also thank Thai Bui and Advanced Nanomaterials for sharing details behind their process for synthesizing endofullerenes in bulk. The authors thank Diksha Dhawan, Jay Soni, Ben Lau, and Juan Arrazola for providing access to and guidance on Xanadu’s PennyLane resource estimator.

\newpage 
\section*{References}
\bibliographystyle{iopart-num}
\bibliography{references}

@misc{AdvancedNanomaterials,
  author       = {{Advanced Nanomaterials}},
  howpublished = {URL \url{https://advnanomaterials.com/}},
  note         = {Accessed: 2026-01-22}
}

@article{Malz2024,
	author = {D. Malz and G. Styliaris and Z.--Y. Wei and J. I Cirac},
	journal = {Physical Review Letters},
	pages = {040404},
	title = {{Preparation of Matrix Product States with Log--Depth Quantum Circuits}},
	volume = {132},
    doi = {10.1103/PhysRevLett.132.040404},
	year = {2024}}

@article{Onuki2019,
	author = {A. Onuki},
	journal = {The Journal of Chemical Physics},
	pages = {134118},
	title = {{Theory of applying heat flow from thermostatted boundary walls: Dissipative and local--equilibrium responses and fluctuation theorems}},
    doi = {10.1063/1.5110877},
	volume = {151},
	year = {20199}}

@article{Litinski2019,
	author = {D. Litinski},
	journal = {Quantum},
	pages = {128},
	title = {{A Game of Surface Codes}},
    doi = {10.22331/q-2019-03-05-128},
	volume = {3},
	year = {2019}}

@article{Horsman2012,
	author = {D. Horsman and A. G. Fowler and S. Devitt and R. {Van Meter}},
	journal = {New Journal of Physics},
	pages = {123011},
	title = {{Surface code quantum computing by lattice surgery}},
    doi = {10.1088/1367-2630/14/12/123011},
	volume = {14},
	year = {2012}}

@article{Poulin2018,
	author = {D. Poulin and A. Kitaev and D. S. Steiger and M. B. Hastings and M. Troyer},
	journal = {Physical Review Letters},
	pages = {010501},
	title = {{Quantum Algorithm for Spectral Measurement with a Lower Gate Count}},
	volume = {121},
    doi = {10.1103/PhysRevLett.121.010501},
	year = {2018}}

@article{Berry2018,
	author = {D. W. Berry and M. Kieferov\'{a} and A. Scherer and Y. R. Sanders and G. H. Low and N. Wiebe and C. Gidney and R. Babbush},
	journal = {NPJ Quantum Inf.},
	pages = {22},
	title = {{Improved techniques for preparing eigenstates of fermionic Hamiltonians}},
    doi ={10.1038/s41534-018-0071-5},
	volume = {4},
	year = {2018}}

@article{Cohn2021,
	author = {J. Cohn and M. Motta and R. M. Parrish},
	journal = {PRX Quantum},
	pages = {040352},
	title = {{Quantum Filter Diagonalization with Compressed Double--Factorized Hamiltonians}},
    doi ={10.1103/PRXQuantum.2.040352},
	volume = {2},
	year = {2021}}

@article{Berry2015,
	author = {D. W. Berry and A. M. Childs and R. Cleve and R. Kothari and R. D. Somma},
	journal = {Physical Review Letters},
	pages = {090502},
	title = {{Simulating Hamiltonian Dynamics with a Truncated Taylor Series}},
    doi = {10.1103/PhysRevLett.114.090502},
	volume = {114},
	year = {2015}}

@article{Childs2012,
	author = {A. M. Childs and N. Wiebe},
	journal = {Quantum Inf. Comput.},
	pages = {901},
	title = {{Hamiltonian simulation using linear combinations of unitary operations}},
	volume = {12},
    doi = {10.26421/QIC12.11-12},
	year = {2012}}

@article{Kuhne2007,
	author = {T. D. Kuhne and M. Krack and F. R. Mohamed and M. Parrinello},
	journal = {Physical Review Letters},
	pages = {066401},
	title = {{Efficient and accurate Car--Parrinello--like approach to Born--Oppenheimer molecular dynamics}},
    doi = {10.1103/PhysRevLett.98.066401},
	volume = {98},
	year = {2007}}

@article{Ricci2003,
	author = {A. Ricci and G. Ciccotti},
	journal = {Molecular Physics},
	pages = {1927-1931},
	title = {{Algorithms for Brownian dynamics}},
	volume = {101},
    doi = {10.1080/0026897031000108113},
	year = {2003}}

@article{Grimme2017,
	author = {S. Grimme and C. Bannwarth and P. Shushkov},
	journal = {J. Chem. Theory. Comput.},
	pages = {1989},
	title = {{A Robust and Accurate Tight--Binding Quantum Chemical Method for Structures, Vibrational Frequencies, and Noncovalent Interactions of Large Molecular Systems Parametrized for All spd--Block Elements (Z = 1-86)}},
	volume = {13},
    doi ={10.1021/acs.jctc.7b00118},
	year = {2017}}

@incollection{Jonsson1998,
	author = {H. J\'{o}nsson and G. Mills and K. W. Jacobsen},
	booktitle = {Classical and Quantum Dynamics in Condensed Phase Systems},
	editor = {B. J. Berne and G. Cicotti and D. F. Coker},
	publisher = {World Scientific},
	year = {1998}}

@article{Henkelman2000b,
	author = {G. Henkelman and H. J\'{o}nsson},
	journal = {The Journal of Chemical Physics },
	pages = {9978-9985 },
	title = {{Improved tangent estimate in the nudged elastic band method for finding minimum energy paths and saddle points}},
    doi = {10.1063/1.1323224},
	volume = {113},
	year = {2000}}

@article{Henkelman2000a,
	author = {G. Henkelman and H. J\'{o}nsson},
	journal = {The Journal of Chemical Physics},
	pages = {9901-9904},
	title = {{A climbing image nudged elastic band method for finding saddle points and minimum energy paths}},
    doi = {10.1063/1.1329672},
	volume = {113},
	year = {2000}}

@article{Grimme2010,
	author = {S. Grimme and J. Antony and S. Ehrlich and H. Kreig},
	journal = {The Journal of Chemical Physics},
	pages = {154104},
	title = {{A consistent and accurate ab initio parametrization of density functional dispersion correction (DFT--D) for the 94 elements H-Pu }},
	volume = {132},
    doi = {10.1063/1.3382344},
	year = {2010}}

@article{Perdew1996,
	author = {P. Perdew and K. Burke and M. Ernzerhof},
	journal = {Physical Review Letters},
	pages = {3865-3868},
	title = {{Generalized Gradient Approximation Made Simple}},
    doi ={10.1103/PhysRevLett.77.3865},
	volume = {77},
	year = {1996}}

@article{Goedecker1996,
	author = {S. Goedecker and M. Teter and J. Hutter},
	journal = {Physical Review E},
	pages = {1703},
	title = {{Separable dual--space Gaussian pseudopotentials}},
	volume = {54},
	year = {1996}}

@article{Schaefer1994,
	author = {A. Schaefer and C. Huber and R. Ahlrichs},
	journal = { The Journal of Chemical Physics},
	pages = {5829-5835},
	title = {{Fully optimized contracted Gaussian--basis sets of triple zeta valence quality for atoms Li to Kr}},
	volume = {100},
    doi = {0.1063/1.467146},
	year = {1994}}

@article{VTST_for_rxn_pathway,
	author = {Hu, Wei-Ping and Liu, Yi-Ping and Truhlar, Donald G.},
	doi = {10.1039/ft9949001715},
	issn = {1364-5455},
	journal = {Journal of the Chemical Society, Faraday Transactions},
	number = {12},
	pages = {1715},
	publisher = {Royal Society of Chemistry (RSC)},
	title = {Variational transition-state theory and semiclassical tunnelling calculations with interpolated corrections: a new approach to interfacing electronic structure theory and dynamics for organic reactions},
	url = {http://dx.doi.org/10.1039/FT9949001715},
	volume = {90},
	year = {1994}}

@article{sun2018pyscf,
	author = {Sun, Qiming and Berkelbach, Timothy C and Blunt, Nick S and Booth, George H and Guo, Sheng and Li, Zhendong and Liu, Junzi and McClain, James D and Sayfutyarova, Elvira R and Sharma, Sandeep and others},
	journal = {Wiley Interdisciplinary Reviews: Computational Molecular Science},
	number = {1},
	pages = {e1340},
	publisher = {Wiley Online Library},
	title = {{PySCF}: the {Python}-based simulations of chemistry framework},
	volume = {8},
    doi = {10.1002/wcms.1340},
	year = {2018}}

@article{fowler2012surface,
	author = {Fowler, Austin G and Mariantoni, Matteo and Martinis, John M and Cleland, Andrew N},
	doi = {10.1103/PhysRevA.86.032324},
	journal = {Physical Review A},
	number = {3},
	pages = {032324},
	publisher = {APS},
	title = {Surface codes: Towards practical large-scale quantum computation},
	volume = {86},
	year = {2012}}

@article{bravyi2005universal,
	author = {Bravyi, Sergey and Kitaev, Alexei},
	journal = {Physical Review A},
	number = {2},
	pages = {022316},
	publisher = {APS},
	title = {Universal quantum computation with ideal {Clifford} gates and noisy ancillas},
	volume = {71},
    doi = {10.1103/PhysRevA.71.022316},
	year = {2005}}

@article{dunning1989gaussian,
	author = {Dunning Jr, Thom H},
	journal = {The Journal of Chemical Physics},
	number = {2},
	pages = {1007--1023},
	publisher = {American Institute of Physics},
	title = {Gaussian basis sets for use in correlated molecular calculations. I. The atoms boron through neon and hydrogen},
	volume = {90},
    doi ={10.1063/1.456153},
	year = {1989}}

@article{von2021quantum,
	author = {von Burg, Vera and Low, Guang Hao and H{\"a}ner, Thomas and Steiger, Damian S and Reiher, Markus and Roetteler, Martin and Troyer, Matthias},
	journal = {Physical Review Research},
	number = {3},
	pages = {033055},
	publisher = {APS},
	title = {Quantum computing enhanced computational catalysis},
    doi = {10.1103/PhysRevResearch.3.033055},
	volume = {3},
	year = {2021}}

@article{hehre1969self,
	author = {Hehre, Warren J and Stewart, Robert F and Pople, John A},
	journal = {The Journal of Chemical Physics},
	number = {6},
	pages = {2657--2664},
	publisher = {American Institute of Physics},
	title = {Self-consistent molecular-orbital methods. I. Use of Gaussian expansions of Slater-type atomic orbitals},
	volume = {51},
    doi = {10.1063/1.1672392},
	year = {1969}}

@misc{beverland2022assessing,
	archiveprefix = {arXiv},
	author = {Beverland, Michael E and Murali, Prakash and Troyer, Matthias and Svore, Krysta M and Hoefler, Torsten and Kliuchnikov, Vadym and Low, Guang Hao and Soeken, Mathias and Sundaram, Aarthi and Vaschillo, Alexander},
	eprint = {2211.07629},
	title = {Assessing requirements to scale to practical quantum advantage},
	year = {2022}}

@misc{Microsoft_Azure_Quantum_Development,
	author = {Microsoft},
	howpublished = {[Online]},
	title = {Azure Quantum Resource Estimation samples},
	url = {https://github.com/microsoft/qsharp/tree/main/samp les/estimation},
	year = {2024}}

@inproceedings{Azure_Quantum_Resource_Estimator,
	address = {New York, NY, USA},
	author = {van Dam, Wim and Mykhailova, Mariia and Soeken, Mathias},
	booktitle = {Proceedings of the SC '23 Workshops of The International Conference on High Performance Computing, Network, Storage, and Analysis},
	doi = {10.1145/3624062.3624211},
	isbn = {9798400707858},
	numpages = {6},
	pages = {1414--1419},
	publisher = {Association for Computing Machinery},
	series = {SC-W '23},
	title = {{Using Azure Quantum Resource Estimator for Assessing Performance of Fault Tolerant Quantum Computation}},
	url = {https://doi.org/10.1145/3624062.3624211},
	year = {2023}}

@inproceedings{Gharibian2022,
	author = {Gharibian, Sevag and Le Gall, Fran\c{c}ois},
	booktitle = {Proceedings of the 54th Annual ACM SIGACT Symposium on Theory of Computing},
	collection = {STOC '22},
	doi = {10.1145/3519935.3519991},
	month = jun,
	publisher = {ACM},
	series = {STOC '22},
	title = {Dequantizing the Quantum singular value transformation: hardness and applications to Quantum chemistry and the Quantum PCP conjecture},
	url = {http://dx.doi.org/10.1145/3519935.3519991},
	year = {2022}}

@article{Low2019hamiltonian,
	author = {Low, Guang Hao and Chuang, Isaac L.},
	doi = {10.22331/q-2019-07-12-163},
	issn = {2521-327X},
	journal = {{Quantum}},
	month = jul,
	pages = {163},
	publisher = {{Verein zur F{\"{o}}rderung des Open Access Publizierens in den Quantenwissenschaften}},
	title = {Hamiltonian {S}imulation by {Q}ubitization},
	url = {https://doi.org/10.22331/q-2019-07-12-163},
	volume = {3},
	year = {2019}}

@article{cp2k,
	author = {K\"{u}hne, Thomas D. and Iannuzzi, Marcella and Del Ben, Mauro and Rybkin, Vladimir V. and Seewald, Patrick and Stein, Frederick and Laino, Teodoro and Khaliullin, Rustam Z. and Sch\"{u}tt, Ole and Schiffmann, Florian and Golze, Dorothea and Wilhelm, Jan and Chulkov, Sergey and Bani-Hashemian, Mohammad Hossein and Weber, Val{\'e}ry and Bor{\v s}tnik, Urban and Taillefumier, Mathieu and Jakobovits, Alice Shoshana and Lazzaro, Alfio and Pabst, Hans and M\"{u}ller, Tiziano and Schade, Robert and Guidon, Manuel and Andermatt, Samuel and Holmberg, Nico and Schenter, Gregory K. and Hehn, Anna and Bussy, Augustin and Belleflamme, Fabian and Tabacchi, Gloria and Gl\"{o}{\ss}, Andreas and Lass, Michael and Bethune, Iain and Mundy, Christopher J. and Plessl, Christian and Watkins, Matt and VandeVondele, Joost and Krack, Matthias and Hutter, J\"{u}rg},
	doi = {10.1063/5.0007045},
	issn = {1089-7690},
	journal = {The Journal of Chemical Physics},
    pages = {0007045},
	month = may,
	number = {19},
	publisher = {AIP Publishing},
	title = {CP2K: An electronic structure and molecular dynamics software package - Quickstep: Efficient and accurate electronic structure calculations},
	url = {http://dx.doi.org/10.1063/5.0007045},
	volume = {152},
	year = {2020}}

@article{Chen2011,
	author = {Chen, Jien-Lian and Hu, Wei-Ping},
	doi = {10.1021/ja203428x},
	issn = {1520-5126},
	journal = {Journal of the American Chemical Society},
	month = sep,
	number = {40},
	pages = {16045--16053},
	publisher = {American Chemical Society (ACS)},
	title = {Theoretical Prediction on the Thermal Stability of Cyclic Ozone and Strong Oxygen Tunneling},
	url = {http://dx.doi.org/10.1021/ja203428x},
	volume = {133},
	year = {2011}}

@article{Buenker1974,
	author = {Buenker, Robert J. and Peyerimhoff, Sigrid D.},
	doi = {10.1007/bf02394557},
	issn = {1432-2234},
	journal = {Theoretica Chimica Acta},
	month = aug,
	number = {1},
	pages = {33--58},
	publisher = {Springer Science and Business Media LLC},
	title = {Individualized configuration selection in CI calculations with subsequent energy extrapolation},
	url = {http://dx.doi.org/10.1007/BF02394557},
	volume = {35},
	year = {1974}}

@article{Buenker1978,
	author = {Buenker, Robert J. and Peyerimhoff, Sigrid D. and Butscher, Werner},
	doi = {10.1080/00268977800100581},
	issn = {1362-3028},
	journal = {Molecular Physics},
	month = mar,
	number = {3},
	pages = {771--791},
	publisher = {Informa UK Limited},
	title = {Applicability of the multi-reference double-excitation CI (MRD-CI) method to the calculation of electronic wavefunctions and comparison with related techniques},
	url = {http://dx.doi.org/10.1080/00268977800100581},
	volume = {35},
	year = {1978}}

@article{McArdle2020,
	author = {McArdle, Sam and Endo, Suguru and Aspuru-Guzik, Al{\'a}n and Benjamin, Simon C. and Yuan, Xiao},
	doi = {10.1103/revmodphys.92.015003},
	issn = {1539-0756},
	journal = {Reviews of Modern Physics},
	month = mar,
	number = {1},
	pages = {015003},
	publisher = {American Physical Society (APS)},
	title = {Quantum computational chemistry},
	url = {http://dx.doi.org/10.1103/RevModPhys.92.015003},
	volume = {92},
	year = {2020}}

@article{Werner1988,
	author = {Werner, Hans-Joachim and Knowles, Peter J.},
	doi = {10.1063/1.455556},
	issn = {1089-7690},
	journal = {The Journal of Chemical Physics},
	month = nov,
	number = {9},
	pages = {5803--5814},
	publisher = {AIP Publishing},
	title = {An efficient internally contracted multiconfiguration--reference configuration interaction method},
	url = {http://dx.doi.org/10.1063/1.455556},
	volume = {89},
	year = {1988}}

@article{linear_t_complexity_dpw,
	author = {Babbush, Ryan and Gidney, Craig and Berry, Dominic W. and Wiebe, Nathan and McClean, Jarrod and Paler, Alexandru and Fowler, Austin and Neven, Hartmut},
	doi = {10.1103/physrevx.8.041015},
	issn = {2160-3308},
	journal = {Physical Review X},
	month = oct,
	number = {4},
	publisher = {American Physical Society (APS)},
	title = {Encoding Electronic Spectra in Quantum Circuits with Linear T Complexity},
	url = {http://dx.doi.org/10.1103/PhysRevX.8.041015},
	volume = {8},
	year = {2018}}

@article{first_quantized_pw,
	author = {Su, Yuan and Berry, Dominic W. and Wiebe, Nathan and Rubin, Nicholas and Babbush, Ryan},
	doi = {10.1103/prxquantum.2.040332},
	issn = {2691-3399},
	journal = {PRX Quantum},
	month = nov,
	number = {4},
	publisher = {American Physical Society (APS)},
	title = {Fault-Tolerant Quantum Simulations of Chemistry in {First Quantization}},
	url = {http://dx.doi.org/10.1103/PRXQuantum.2.040332},
    pages = {040332},
	volume = {2},
	year = {2021}}

@article{Kroto1985,
	author = {Kroto, H. W. and Heath, J. R. and O'Brien, S. C. and Curl, R. F. and Smalley, R. E.},
	doi = {10.1038/318162a0},
	issn = {1476-4687},
	journal = {Nature},
	month = nov,
	number = {6042},
	pages = {162--163},
	publisher = {Springer Science and Business Media LLC},
	title = {C60: Buckminsterfullerene},
	url = {http://dx.doi.org/10.1038/318162a0},
	volume = {318},
	year = {1985}}

@article{Saunders1993,
	author = {Saunders, Martin and Jim{\'e}nez-V{\'a}zquez, Hugo A. and Cross, R. James and Poreda, Robert J.},
	doi = {10.1126/science.259.5100.1428},
	issn = {1095-9203},
	journal = {Science},
	month = mar,
	number = {5100},
	pages = {1428--1430},
	publisher = {American Association for the Advancement of Science (AAAS)},
	title = {Stable Compounds of Helium and Neon: He@C 60 and Ne@C 60},
	url = {http://dx.doi.org/10.1126/science.259.5100.1428},
	volume = {259},
	year = {1993}}

@article{hydrogen_storage,
	author = {Pupysheva, Olga V. and Farajian, Amir A. and Yakobson, Boris I.},
	doi = {10.1021/nl071436g},
	issn = {1530-6992},
	journal = {Nano Letters},
	month = oct,
	number = {3},
	pages = {767--774},
	publisher = {American Chemical Society (ACS)},
	title = {Fullerene Nanocage Capacity for Hydrogen Storage},
	url = {http://dx.doi.org/10.1021/nl071436g},
	volume = {8},
	year = {2007}}

@article{vtst_review_paper,
	author = {Bao, Junwei Lucas and Truhlar, Donald G.},
	doi = {10.1039/c7cs00602k},
	issn = {1460-4744},
	journal = {Chemical Society Reviews},
	number = {24},
	pages = {7548--7596},
	publisher = {Royal Society of Chemistry (RSC)},
	title = {Variational transition state theory: theoretical framework and recent developments},
	url = {http://dx.doi.org/10.1039/c7cs00602k},
	volume = {46},
	year = {2017}}

@article{casscf_ozone_study,
	author = {Theis, Daniel and Ivanic, Joseph and Windus, Theresa L. and Ruedenberg, Klaus},
	doi = {10.1063/1.4942019},
	issn = {1089-7690},
	journal = {The Journal of Chemical Physics},
	month = mar,
	number = {10},
	title = {The transition from the open minimum to the ring minimum on the ground state and on the lowest excited state of like symmetry in ozone: A configuration interaction study},
	volume = {144},
    pages = {104304},
	year = {2016}}

@article{thermal_stability_of_cyclic_ozone,
	author = {Chen, Jien-Lian and Hu, Wei-Ping},
	doi = {10.1021/ja203428x},
	issn = {1520-5126},
	journal = {Journal of the American Chemical Society},
	month = sep,
	number = {40},
	pages = {16045--16053},
	publisher = {American Chemical Society (ACS)},
	title = {Theoretical Prediction on the Thermal Stability of Cyclic Ozone and Strong Oxygen Tunneling},
	url = {http://dx.doi.org/10.1021/ja203428x},
	volume = {133},
	year = {2011}}

@article{otten2023qrechem,
	author = {Otten, Matthew and Kang, Byeol and Fedorov, Dmitry and Lee, Joo-Hyoung and Benali, Anouar and Habib, Salman and Gray, Stephen K and Alexeev, Yuri},
	journal = {Frontiers in Quantum Science and Technology},
	pages = {1232624},
	publisher = {Frontiers Media SA},
	title = {QREChem: quantum resource estimation software for chemistry applications},
	volume = {2},
	year = {2023}}

@article{synthesis_of_endofullerenes,
	author = {Bloodworth, Sally and Whitby, Richard J.},
	doi = {10.1038/s42004-022-00738-9},
	issn = {2399-3669},
	journal = {Communications Chemistry},
	month = oct,
	number = {1},
	publisher = {Springer Science and Business Media LLC},
	title = {Synthesis of endohedral fullerenes by molecular surgery},
	url = {http://dx.doi.org/10.1038/s42004-022-00738-9},
    pages={121}, 
	volume = {5},
	year = {2022}}

@article{kitaev1995quantum,
	author = {Kitaev, A Yu},
	archiveprefix = {arXiv},
    journal = {arXiv e-prints},
    eprint = {quant-ph/9511026},
	title = {Quantum measurements and the Abelian stabilizer problem},
	year = {1995}}

@article{entropy_of_ozone,
	author = {Denis Sh. Sabirov and Igor S. Shepelevich},
	doi = {https://doi.org/10.1016/j.comptc.2015.09.016},
	issn = {2210-271X},
	journal = {Computational and Theoretical Chemistry},
	pages = {61-66},
	title = {Information entropy of oxygen allotropes. A still open discussion about the closed form of ozone},
	url = {https://www.sciencedirect.com/science/article/pii/S2210271X15003813},
	volume = {1073},
	year = {2015}}

@article{SHCI_ozone_pes,
	author = {Chien, Alan D. and Holmes, Adam A. and Otten, Matthew and Umrigar, C. J. and Sharma, Sandeep and Zimmerman, Paul M.},
	date = {2018/03/15},
	doi = {10.1021/acs.jpca.8b01554},
	isbn = {1089-5639},
	journal = {The Journal of Physical Chemistry A},
	journal1 = {The Journal of Physical Chemistry A},
	journal2 = {J. Phys. Chem. A},
	month = {03},
	number = {10},
	pages = {2714--2722},
	publisher = {American Chemical Society},
	title = {Excited States of Methylene, Polyenes, and Ozone from Heat-Bath Configuration Interaction},
	type = {doi: 10.1021/acs.jpca.8b01554},
	url = {https://doi.org/10.1021/acs.jpca.8b01554},
	volume = {122},
	year = {2018},
	year1 = {2018}}

@misc{selectedci_dmrg,
	archiveprefix = {arXiv},
	author = {Stepan Fomichev and Kasra Hejazi and Modjtaba Shokrian Zini and Matthew Kiser and Joana Fraxanet Morales and Pablo Antonio Moreno Casares and Alain Delgado and Joonsuk Huh and Arne-Christian Voigt and Jonathan E. Mueller and Juan Miguel Arrazola},
	eprint = {2310.18410},
	primaryclass = {quant-ph},
	title = {Initial state preparation for quantum chemistry on quantum computers},
	year = {2023}}

@misc{patent_uv_synth,
	author = {Naseri, Sara and Bock, Marlene},
	month = jan # {~7},
	note = {US Patent 8,623,337},
	publisher = {Google Patents},
	title = {Endohedral fullerenes having enclosed therein one or more ozone molecules, and their use as a UV-absorbing agent},
	year = {2014}}

@article{3zeta_dft_2,
	author = {Sabirov, D.Sh. and Khursan, S.L. and Bulgakov, R.G.},
	doi = {10.1016/j.jmgm.2008.03.006},
	issn = {1093-3263},
	journal = {Journal of Molecular Graphics and Modelling},
	month = sep,
	number = {2},
	pages = {124--130},
	publisher = {Elsevier BV},
	title = {Ozone addition to {C60} and {C70} fullerenes: A {DFT} study},
	url = {http://dx.doi.org/10.1016/j.jmgm.2008.03.006},
	volume = {27},
	year = {2008}}

@article{3zeta_dft_1,
	author = {Sabirov, Denis Sh. and Bulgakov, Ramil G. and Khursan, Sergey L.},
	doi = {10.1016/j.mencom.2010.06.017},
	issn = {0959-9436},
	journal = {Mendeleev Communications},
	month = jul,
	number = {4},
	pages = {231--233},
	publisher = {Elsevier BV},
	title = {Reactivity of carbonyl oxides generated by the ozonolysis of C60 and C70 fullerenes: a chemiluminescence study and quantum-topological analysis},
	url = {http://dx.doi.org/10.1016/j.mencom.2010.06.017},
	volume = {20},
	year = {2010}}

@article{mgo_crystal_ozone,
	author = {Plass, Richard and Egan, Kenneth and Collazo-Davila, Chris and Grozea, Daniel and Landree, Eric and Marks, Laurence D. and Gajdardziska-Josifovska, Marija},
	doi = {10.1103/PhysRevLett.81.4891},
	issue = {22},
	journal = {Physical Review Letters},
	month = {Nov},
	numpages = {0},
	pages = {4891--4894},
	publisher = {American Physical Society},
	title = {Cyclic Ozone Identified in Magnesium Oxide (111) Surface Reconstructions},
	url = {https://link.aps.org/doi/10.1103/PhysRevLett.81.4891},
	volume = {81},
	year = {1998}}

@techreport{o3_laser_pulse_search,
	author = {Levis, Robert J. and Romanov, Dmitri A. and Rabitz, Hersch A. and Kevrekidis, Yannis G. and Coifman, Ronald},
	institution = {TEMPLE UNIV PHILADELPHIA PA},
	month = mar,
	shorttitle = {Photonic {Reagents}},
	title = {Photonic {Reagents}: {The} {Production} of {Cyclic} {Ozone}, {With} a {Focus} on {Developing} {Equation} {Free} {Methods} for {Optimization} {Schemes}},
	url = {https://apps.dtic.mil/sti/citations/ADA480657},
	urldate = {2022-01-11},
	year = {2008}}

@article{fullerene_surgery,
	author = {Murata, Michihisa and Murata, Yasujiro and Komatsu, Koichi},
	doi = {10.1039/B811738A},
	issue = {46},
	journal = {Chemical Communications},
	pages = {6083-6094},
	publisher = {The Royal Society of Chemistry},
	title = {Surgery of fullerenes},
	url = {http://dx.doi.org/10.1039/B811738A},
	year = {2008}}

@misc{laser_search_article,
	journal = {ScienceDaily},
	title = {Temple {Researcher} {Attempting} {To} {Create} {Cyclic} {Ozone}},
	url = {https://www.sciencedaily.com/releases/2005/02/050205122519.htm},
	urldate = {2022-01-11}}

@article{chem_rxn_qc,
	author = {Reiher, Markus and Wiebe, Nathan and Svore, Krysta M. and Wecker, Dave and Troyer, Matthias},
	doi = {10.1073/pnas.1619152114},
	eprint = {https://www.pnas.org/content/114/29/7555.full.pdf},
	issn = {0027-8424},
	journal = {Proceedings of the National Academy of Sciences},
	number = {29},
	pages = {7555--7560},
	publisher = {National Academy of Sciences},
	title = {Elucidating reaction mechanisms on quantum computers},
	url = {https://www.pnas.org/content/114/29/7555},
	volume = {114},
	year = {2017}}

@article{quantum_materials_simulation,
	author = {Babbush, Ryan and Wiebe, Nathan and McClean, Jarrod and McClain, James and Neven, Hartmut and Chan, Garnet Kin-Lic},
	doi = {10.1103/PhysRevX.8.011044},
	issue = {1},
	journal = {Physical Review X},
	month = {Mar},
	numpages = {40},
	pages = {011044},
	publisher = {American Physical Society},
	title = {Low-Depth Quantum Simulation of Materials},
	url = {https://link.aps.org/doi/10.1103/PhysRevX.8.011044},
	volume = {8},
	year = {2018}}

@article{speedup_qchem,
	author = {Lee, Seunghoon and Lee, Joonho and Zhai, Huanchen and Tong, Yu and Dalzell, Alexander M. and Kumar, Ashutosh and Helms, Phillip and Gray, Johnnie and Cui, Zhi-Hao and Liu, Wenyuan and Kastoryano, Michael and Babbush, Ryan and Preskill, John and Reichman, David R. and Campbell, Earl T. and Valeev, Edward F. and Lin, Lin and Chan, Garnet Kin-Lic},
	doi = {10.1038/s41467-023-37587-6},
	issn = {2041-1723},
	journal = {Nature Communications},
	month = apr,
	pages = {1},
	publisher = {Springer Science and Business Media LLC},
	title = {Evaluating the evidence for exponential quantum advantage in ground-state quantum chemistry},
	url = {http://dx.doi.org/10.1038/s41467-023-37587-6},
	volume = {14},
	year = {2023}}

@article{Reiher_femoco,
	author = {Reiher, Markus and Wiebe, Nathan and Svore, Krysta M. and Wecker, Dave and Troyer, Matthias},
	doi = {10.1073/pnas.1619152114},
	issn = {1091-6490},
	journal = {Proceedings of the National Academy of Sciences},
	month = jul,
	number = {29},
	pages = {7555--7560},
	publisher = {Proceedings of the National Academy of Sciences},
	title = {Elucidating reaction mechanisms on quantum computers},
	url = {http://dx.doi.org/10.1073/pnas.1619152114},
	volume = {114},
	year = {2017}}

@article{goings2022reliably,
	author = {Goings, Joshua J and White, Alec and Lee, Joonho and Tautermann, Christofer S and Degroote, Matthias and Gidney, Craig and Shiozaki, Toru and Babbush, Ryan and Rubin, Nicholas C},
	journal = {Proceedings of the National Academy of Sciences},
	number = {38},
	pages = {e2203533119},
	publisher = {National Acad Sciences},
	title = {Reliably assessing the electronic structure of cytochrome P450 on today's classical computers and tomorrow's quantum computers},
    doi ={10.1073/pnas.2203533119},
	volume = {119},
	year = {2022}}

@article{AVAS,
	author = {Sayfutyarova, Elvira R. and Sun, Qiming and Chan, Garnet Kin-Lic and Knizia, Gerald},
	date = {2017/09/12},
	doi = {10.1021/acs.jctc.7b00128},
	isbn = {1549-9618},
	journal = {Journal of Chemical Theory and Computation},
	journal1 = {Journal of Chemical Theory and Computation},
	journal2 = {Journal of Chemical Theory and Computation},
	month = {09},
	number = {9},
	pages = {4063--4078},
	publisher = {American Chemical Society},
	title = {Automated Construction of Molecular Active Spaces from Atomic Valence Orbitals},
	type = {doi: 10.1021/acs.jctc.7b00128},
	url = {https://doi.org/10.1021/acs.jctc.7b00128},
	volume = {13},
	year = {2017},
	year1 = {2017}}

@article{Chapman2020,
	author = {Chapman, Adrian and Flammia, Steven T.},
	doi = {10.22331/q-2020-06-04-278},
	issn = {2521-327X},
	journal = {Quantum},
	month = jun,
	pages = {278},
	publisher = {Verein zur Forderung des Open Access Publizierens in den Quantenwissenschaften},
	title = {Characterization of solvable spin models via graph invariants},
	url = {http://dx.doi.org/10.22331/q-2020-06-04-278},
	volume = {4},
	year = {2020}}

@article{Elman2021,
	author = {Elman, Samuel J. and Chapman, Adrian and Flammia, Steven T.},
	doi = {10.1007/s00220-021-04220-w},
	issn = {1432-0916},
	journal = {Communications in Mathematical Physics},
	month = oct,
	number = {2},
	pages = {969--1003},
	publisher = {Springer Science and Business Media LLC},
	title = {Free Fermions Behind the Disguise},
	url = {http://dx.doi.org/10.1007/s00220-021-04220-w},
	volume = {388},
	year = {2021}}

@article{chapman2023unified,
	archiveprefix = {arXiv},
	author = {Chapman, Adrian and Elman, Samuel J and Mann, Ryan L},
	eprint = {2305.15625},
	journal = {arXiv e-prints},
	title = {A Unified Graph-Theoretic Framework for Free-Fermion Solvability},
	year = {2023}}

@misc{pyliqtr,
	author = {Obenland, K. and Elenewski, J. and Morrell, K. J. and Neumann, R. S. and Kurlej, A. and Rood, R. and Blue, J. and Belarge, J. and Rempfer, B. and Kuklinski, P.},
	copyright = {Creative Commons Attribution 4.0 International},
	doi = {10.5281/ZENODO.10913397},
	publisher = {Zenodo},
	title = {isi-usc-edu/pyLIQTR: Release 1.1.1},
	url = {https://zenodo.org/doi/10.5281/zenodo.10913397},
	year = {2024}}

@article{Ortiz2000QuantumAF,
	author = {Gerardo Ortiz and James E. Gubernatis and Emanuel Knill and Raymond Laflamme},
	doi = {10.1103/PhysRevA.64.022319},
	journal = {Physical Review A},
	pages = {022319},
	title = {Quantum algorithms for fermionic simulations},
	volume = {64},
	year = {2000}}

@article{Kassal2008PolynomialtimeQA,
	author = {Ivan Kassal and Stephen P. Jordan and Peter J. Love and Masoud Mohseni and Al{\'a}n Aspuru‐Guzik},
	doi = {10.1073/pnas.0808245105},
	journal = {Proceedings of the National Academy of Sciences},
	pages = {18681 - 18686},
	title = {Polynomial-time quantum algorithm for the simulation of chemical dynamics},
	volume = {105},
	year = {2008}}

@article{Lanyon2009TowardsQC,
	author = {Ben P. Lanyon and James Daniel Whitfield and Geoff Gillett and Michael E. Goggin and Marcelo P. Almeida and Ivan Kassal and Jacob D. Biamonte and Masoud Mohseni and Benjamin J Powell and Marco Barbieri and Al{\'a}n Aspuru-Guzik and Andrew G. White},
	doi = {10.1038/nchem.483},
	journal = {Nature chemistry},
	pages = {106-11},
	title = {Towards quantum chemistry on a quantum computer},
	volume = {2 2},
	year = {2009}}

@article{Bauer_2020,
	author = {Bauer, Bela and Bravyi, Sergey and Motta, Mario and Chan, Garnet Kin-Lic},
	doi = {10.1021/acs.chemrev.9b00829},
	journal = {Chemical Reviews},
	number = {22},
	pages = {12685-12717},
	title = {Quantum Algorithms for Quantum Chemistry and Quantum Materials Science},
	volume = {120},
	year = {2020}}

@misc{Low2018HamiltonianSI,
	archiveprefix = {arXiv},
	author = {Guang Hao Low and Nathan Wiebe},
	eprint = {1805.00675},
	title = {Hamiltonian Simulation in the Interaction Picture},
	year = {2018}}

@article{hypercontraction,
	author = {Lee, Joonho and Berry, Dominic W. and Gidney, Craig and Huggins, William J. and McClean, Jarrod R. and Wiebe, Nathan and Babbush, Ryan},
	doi = {10.1103/PRXQuantum.2.030305},
	issue = {3},
	journal = {PRX Quantum},
	numpages = {62},
	pages = {030305},
	publisher = {American Physical Society},
	title = {Even More Efficient Quantum Computations of Chemistry Through Tensor Hypercontraction},
	volume = {2},
	year = {2021}}

@article{babbush2018encoding,
	author = {Babbush, Ryan and Gidney, Craig and Berry, Dominic W and Wiebe, Nathan and McClean, Jarrod and Paler, Alexandru and Fowler, Austin and Neven, Hartmut},
	doi = {10.1103/PhysRevX.8.041015},
	journal = {Physical Review X},
	number = {4},
	pages = {041015},
	publisher = {APS},
	title = {Encoding electronic spectra in quantum circuits with linear {T} complexity},
	volume = {8},
	year = {2018},
	pages = {041015}}

@misc{Berry2024DoublingEO,
	archiveprefix = {arXiv},
	author = {Dominic W. Berry and Danial Motlagh and Giacomo Pantaleoni and Nathan Wiebe},
	eprint = {2401.10321},
	title = {Doubling Efficiency of Hamiltonian Simulation via Generalized Quantum Signal Processing},
	year = {2024}}

@article{Motta2021,
	Author = {Motta, Mario and Ye, Erika and McClean, Jarrod R. and Li, Zhendong and Minnich, Austin J. and Babbush, Ryan and Chan, Garnet Kin-Lic},
	Da = {2021/05/27},
	Doi = {10.1038/s41534-021-00416-z},
	Isbn = {2056-6387},
	Journal = {npj Quantum Information},
	Number = {1},
	Pages = {83},
	Title = {Low rank representations for quantum simulation of electronic structure},
	Url = {https://doi.org/10.1038/s41534-021-00416-z},
	Volume = {7},
	Year = {2021}}

@article{nature2025,
  ISSN = {1755-4349},
  url = {http://dx.doi.org/10.1038/s41557-025-01919-4},
  DOI = {10.1038/s41557-025-01919-4},
  journal = {Nature Chemistry},
  publisher = {Springer Science and Business Media LLC},
  year = {2025},
  month = aug 
}

@article{Sharma2025,
  title = {Encapsulation effects on the structure and stability of O3,  SO2,  S2O,  and S3 in C60,  C70,  and C80 fullerenes.},
  volume = {1245},
  ISSN = {2210-271X},
  url = {http://dx.doi.org/10.1016/j.comptc.2025.115096},
  DOI = {10.1016/j.comptc.2025.115096},
  journal = {Computational and Theoretical Chemistry},
  publisher = {Elsevier BV},
  author = {Sharma,  Vishal and Nagpal,  Vasu and Chakraborty,  Aniruddha},
  year = {2025},
  month = mar,
  pages = {115096}
}

@misc{camps2025quantumcomputingtechnologyroadmaps,
      title={Quantum Computing Technology Roadmaps and Capability Assessment for Scientific Computing -- An analysis of use cases from the NERSC workload}, 
      author={Daan Camps and Ermal Rrapaj and Katherine Klymko and Hyeongjin Kim and Kevin Gott and Siva Darbha and Jan Balewski and Brian Austin and Nicholas J. Wright},
      year={2025},
      eprint={2509.09882},
      archivePrefix={arXiv},
      primaryClass={quant-ph},
      url={https://arxiv.org/abs/2509.09882}, 
}

@article{Low_2025,
   title={Fast Quantum Simulation of Electronic Structure by Spectral Amplification},
   volume={15},
   ISSN={2160-3308},
   url={http://dx.doi.org/10.1103/pb2g-j9cw},
   DOI={10.1103/pb2g-j9cw},
   number={4},
   journal={Physical Review X},
   publisher={American Physical Society (APS)},
   author={Low, Guang Hao and King, Robbie and Berry, Dominic W. and Han, Qiushi and DePrince, A. Eugene and White, Alec F. and Babbush, Ryan and Somma, Rolando D. and Rubin, Nicholas C.},
   year={2025},
   month=oct }

@article{bergholm2018pennylane,
  title={Pennylane: Automatic differentiation of hybrid quantum-classical computations},
  author={Bergholm, Ville and Izaac, Josh and Schuld, Maria and Gogolin, Christian and Ahmed, Shahnawaz and Ajith, Vishnu and Alam, M Sohaib and Alonso-Linaje, Guillermo and AkashNarayanan, B and Asadi, Ali and others},
  journal={arXiv preprint arXiv:1811.04968},
  year={2018}
}

@misc{JaySoni2026,
    title = "How to use PennyLane for Resource Estimation",
    author = "Jay Soni and Anton Naim Ibrahim",
    year = "2026",
    month = "1",
    journal = "PennyLane Demos",
    publisher = "Xanadu",
    howpublished = "\url{https://pennylane.ai/qml/demos/re_how_to_use_pennylane_for_resource_estimation}",
    note = "Date Accessed: 2026-02-02"
}

@article{Trache2020Nanoenergetic,
  author  = {Trache, Djalal and DeLuca, Luigi T.},
  title   = {Nanoenergetic Materials: Preparation, Properties, and Applications},
  journal = {Nanomaterials},
  year    = {2020},
  volume  = {10},
  number  = {12},
  pages   = {2347},
  doi     = {10.3390/nano10122347}
}

@article{Thiruvengadathan2024Nanoenergetic,
  author  = {Thiruvengadathan, Rajagopalan and Wang, Anqi},
  title   = {Nanoenergetic Materials: From Materials to Applications},
  journal = {Nanomaterials},
  year    = {2024},
  volume  = {14},
  number  = {19},
  pages   = {1574},
  doi     = {10.3390/nano14191574}
}

@article{Rossi2007MEMS,
  author  = {Rossi, Carole and Zhang, Kaili and Est{\`e}ve, Daniel and Alphonse, Pierre and Tailhades, Philippe and Vahlas, Constantin},
  title   = {Nanoenergetic materials for {MEMS}: A review},
  journal = {Journal of Microelectromechanical Systems},
  year    = {2007},
  volume  = {16},
  number  = {4},
  pages   = {919--931},
  doi     = {10.1109/JMEMS.2007.893519}
}

@article{Martirosyan2011GasGenerators,
  author  = {Martirosyan, Karen S.},
  title   = {Nanoenergetic Gas-Generators: principles and applications},
  journal = {Journal of Materials Chemistry},
  year    = {2011},
  volume  = {21},
  pages   = {9400--9405},
  doi     = {10.1039/C1JM11300C}
}

@article{Popov2013Endohedral,
  author  = {Popov, Alexey A. and Yang, Shangfeng and Dunsch, Lothar},
  title   = {Endohedral fullerenes},
  journal = {Chemical Reviews},
  year    = {2013},
  volume  = {113},
  number  = {8},
  pages   = {5989--6113},
  doi     = {10.1021/cr300297r}
}

@article{Jalife2020NgEndohedral,
  author  = {Jalife, Said and Arcudia, Jessica and Pan, Sudip and Merino, Gabriel},
  title   = {Noble gas endohedral fullerenes},
  journal = {Chemical Science},
  year    = {2020},
  volume  = {11},
  pages   = {6642--6652},
  doi     = {10.1039/D0SC02507K}
}

@article{Xu2024MicrothrusterReview,
  author  = {Xu, Jianbing and Zhang, Jiangtao and Li, Fuwei and Liu, Shiyi and Ye, Yinghua and Shen, Ruiqi},
  title   = {A review on solid propellant micro-thruster array based on {MEMS} technology},
  journal = {Fuel Processing Technology},
  year    = {2024},
  note    = {Open access review; journal/volume/pages may vary by indexing},
  doi     = {10.1016/j.fpc.2023.03.002}
}

\end{document}